\documentclass[epj]{svjour}
\usepackage{graphicx}
\usepackage{bm}   
\usepackage{color}
\usepackage{amssymb}
\usepackage{amsmath}
\usepackage{balance}
\usepackage{subfigure}
\usepackage{enumitem}
\begin{document}
	
	\title{Mitigation of extreme events in an excitable system}
	\author{\textbf{R. Shashangan}\inst{1}\textbf{,} \textbf{S. Sudharsan}\inst{2}\textbf{,} \textbf{A. Venkatesan}\inst{3}\textbf{,} \textbf{M. Senthilvelan}\inst{1,a}\mail{velan@cnld.bdu.ac.in}
	}                     
	%
	%
	\institute{Department of Nonlinear Dynamics, Bharathidasan University, Tiruchirappalli 620024, Tamil Nadu, India. \and Physics and Applied Mathematics Unit, Indian Statistical Institute, Kolkata 700108, West Bengal, India \and Department of Physics, Nehru Memorial College (Autonomous), Tiruchirappalli 621007, Tamil Nadu, India.}
	\date{Received: date / Revised version: date}
	%
	\abstract{
		Formulating mitigation strategies is one of the main aspect in the dynamical study of extreme events. Apart from the effective control, easy implementation of the devised tool should also be given importance. In this work, we analyze the mitigation of extreme events in a coupled FitzHugh-Nagumo (FHN) neuron model utilizing an easily implementable constant bias analogous to a constant DC stimulant. We report the route through which the extreme events gets mitigated in $Two$, $Three$ and $N-$coupled FHN systems. In all the three cases, extreme events in the observable $\bar{x}$ gets suppressed. We confirm our results with the probability distribution function of peaks, $d_{max}$ plot and probability plots. Here $d_{max}$ is a measure of number of standard deviations that crosses the average amplitude corresponding to $\bar{x}_{max}$. Interestingly, we found that constant bias suppresses the extreme events without changing the collective frequency of the system. 
		\PACS{
			{PACS-key}{Extreme Events\and FitzHugh-Nagumo model\and Constant bias}
		} 
	}
	\titlerunning{Mitigation of extreme events in an excitable system}
	\authorrunning{R. Shashangan et al.}
	\maketitle
	%
\section{Introduction}
\par All physical and physiological functions taking place in animals and humans are due to the complex action of neurons in the brain which occurs as a result of synaptic interconnection between several billions of neurons. There are more number of interactions than the total number of neurons which lead to the real complexity. Each neuron independently is an excitable system and fires the action potential in combination with other neurons when performing an action \cite{r1,r2}. During collective firing, self-organization occurs resulting in several spatiotemporal patterns like incoherent \cite{r3}, chimera \cite{r4}, cluster \cite{r5}, solitary state \cite{r6}, complete coherent state \cite{r7} and so on. Inspite of several manifestations, the dynamics of individual neurons in the collective state turns out to be either periodic or chaotic (bounded chaos with small fluctuations). But under certain conditions, the time series of the system produce an unanticipated large deviations that significantly differs from a regular level \cite{r8,r9}. Such an abnormal deviation in the activity of neurons occurs whenever a disorder called epileptic seizures take place. Seizures are extremely debilitating and some epileptic seizures are characterized by perceivable rhythmic muscle contractions \cite{r10}.  

Although research on these epileptic seizures from the physiological or biological point of view is prevalent, it is important to carry out further research on epileptic seizures from the perspective of nonlinear dynamics. In the context of nonlinear dynamics, epileptic seizures are termed as extreme events. Instabilities in the dynamical systems make these events occasional and intermittent. Existence of these events are not only restricted to neural networks but are ubiquitous in almost all scientific disciplines. Rogue waves \cite{r11}, earthquakes, cyclones, epidemics \cite{r9}, share market crash \cite{r12}, algal blooms \cite{r13}, and power blackouts \cite{r14} are examples of some of the extreme events in real world. Interior crisis, intermittency, alternate inflation and deflation of chaotic attractor, stick and slide near discontinuous boundary, influence of noise, attractor bubbling and transient instabilities \cite{r15,r16,r17,r18,r19,r20,r21} are some of the potential mechanisms reported as the causes for the emanation of extreme events in various dynamical systems such as  Li\'enard \cite{r22}, Hindmarsh-Rose (HR) model \cite{r23}, Micro-Electromechanical Systems (MEMS) \cite{r24}, FitzHugh-Nagumo (FHN) model \cite{r8}, Mechanical oscillator \cite{r17}, Optical system \cite{r18} and Duffing systems \cite{r26}. Among these, FHN and HR models represent the time evolution of individual neurons. Extreme events have emerged as a result of the opening of channel-like structure for a $Two-$coupled FHN system and also as a result of excitable proto-events for a $N-$coupled FHN system \cite{r8}. In the case of HR model, instability in the antiphase synchronization manifold is the reason behind the dragon king like events in the HR model \cite{r23}. In this context, the emergence of extreme events in a network of Josephson junction \cite{r25} and Li\'enard type system \cite{r22} occurs due to the interplay of heterogenous damping parameter along with the coupling strength. In addition to this, extreme events also occur through the variation of potential energy of the system \cite{r26}, pulse-shaped explosion \cite{r27}, interplay between degree distribution and repulsive interaction \cite{r28}, and interplay of stochastic noise with multistable dynamics \cite{r21}. Recently dragon king extreme events have been observed and predicted using machine learning techniques \cite{r29,r30}, and quasiperiodic route to extreme events via strange non-chaotic attractor (SNA)\cite{r31}, while extreme events in non-chaotic regimes \cite{r32} have also been observed.

Although several studies have been dedicated towards determining the mechanism trailing the emanation of extreme events, it is equally important to devise strategies to mitigate these extreme events, specifically in networks. Hence, the need for devising and developing an easily implementable mitigation tool has become a pressing priority. In the past few years, various tools have been used to mitigate extreme events arising in both isolated and coupled dynamical systems, such as threshold activated coupling \cite{r16}, external forcing \cite{r17}, time-delayed feedback \cite{r33,r34}, noise \cite{r35}, localized perturbation \cite{r36} and network mobiling \cite{r37}. But there is no unique mitigation method which is applicable to all systems. Every mitigation method has its own merits and demerits. So mitigation methods are system specific. One more important factor that one has to take care of while mitigating extreme events in networks is to ensure that the technique should mitigate only extreme events and it should not change the collective dynamics of the system. Moreover, with all the previous procedures being feedback type, it is difficult to implement them to control extreme events. So similar to isolated systems, it is crucial to devise procedures to mitigate extreme events in networks using a technique that is easy to implement.    

In this context, we intend to use an easily implementable tool to mitigate extreme events arising in a network representing coupled neurons. For this purpose, we utilize a simple but a powerful nonlinear model, namely FHN model, which is simple in its form but effectively captures all the properties of a neuron. For this model, the authors in Refs. \cite{r8} and \cite{r15} have discussed the emergence of extreme events and in Ref. \cite{r38}, the authors found that extreme events in this system is controlled under environmental coupling. Since the brain is a complex system having a very complicated structure, creating an enviroment and then controlling it is very difficult. Therefore in search of an easily implementable tool for control, recently, in the Ref. \cite{r39}, the authors have found that a constant bias can effectively suppress extreme events in an isolated mechanical system. Since constant bias being a well-known tool in the literature of nonlinear dynamics for its easy implementation, we adapted it in our study of the suppression of extreme events in a network of FHN neurons. On the other hand, in biophysiology, there is a tool called Deep Brain Stimulation (DBS) which is used to control epileptic seizures \cite{r43,r44} as an alternate to medications. Similar to the constant bias, this technique is also easy to implement and monitor. Most importantly DBS has an ample success rate in controlling epilepsy.  So, we can clearly see that constant bias (in dynamical system) serves as a counterpart to DBS (in physiology). In a nutshell, the advantages of using constant bias are:
\begin{enumerate}[label=(\roman*)]
	\item It is a non-feedback method - Changes are being made in the control parameters of the system rather than in the system's position in phase space. Therefore, by incorporating this method, changes are produced in the system but without completely changing the dynamics. The foremost advantage of utilizing the non-feedback technique is that it is fast, high resilient, online monitoring and processing is not necessary. 
	\item Implementation is easy - As far as the implementation is concerned, from applications point of view, constant bias usually represents constant current or constant voltage \cite{r45}. The DC field can be easily implemented in experiments, specifically in electronic experiments as an external source.
\end{enumerate}
\par By considering these factors, in this work, we apply a constant bias to a $Two$, $Three$ and $N-$coupled network of FHN dynamical system and observe how it controls extreme events and thereby determine the dynamics behind the suppression. To the best of our knowledge it is the first time that a constant bias is utilized to control extreme events in a network of FHN neurons. Upon our analysis, we have found that even very weak constant bias of lower magnitude can suppress these extreme events in the considered system. Interestingly, we have found that this very weak constant bias suppresses extreme events without affecting the collective frequency of the system. 
\par Our study is organized in the following manner. In Sec. 2, we define the mathematical model under consideration. In Sec. 3, we list out all the results that we have obtained for $Two$, $Three$ and $N-$coupled neurons and its corresponding mechanism of suppression. In Sec. 4, we discuss in detail about how these events are suppressed in this system using constant bias. We conclude about our work in Sec. 5.
\section{Model and Analysis}
The excitable system considered for our analysis is the FHN model which describes a neuron's excitability \cite{r46}. FHN system is the two dimensional reduction of Hodgkin-Huxley (HH) model. The corresponding equation reads \cite{r1} 
\begin{eqnarray}
	\dot{x}  &=& x(a - x)(x - 1) - y + I, \nonumber \\
	\dot{y}  &=& bx - cy
\end{eqnarray}
\par The state variables $x$ and $y$ corresponds to the membrane potential and recovery variable which is related to the refractory period. The parameter $I$ refers to the stimulus current. The parameter $a$ represents the shape of the cubic parabola, $b$ and $c$ are arbitrary parameters. When the value of $I$ is zero and the value of $b$ and $c$ are small the dynamics exhibited by this model is relaxation oscillation \cite{r1}. 
\par For our analysis, we consider a globally coupled network of FHN model \cite{r8}. The corresponding network equation is given by 
\begin{eqnarray}
	\dot{x_{i}}  &=& x_{i}(a_{i} - x_{i})(x_{i} - 1) - y_{i} + k\sum_{j=1}^{n} A_{ij}(x_{j} - x_{i}) + s_{x}, \nonumber \\
	\dot{y_{i}}  &=& b_{i}x_{i} - c_{i}y_{i} + s_{y}, \quad i = 1, 2, 3,...,N
\end{eqnarray}
where the index $i$ corresponds to the number of neurons. The parameter $k$ is the coupling strength, $s_{x}$ is the constant bias applied to the membrane potential variable $(x)$ and $s_{y}$ is the constant bias applied to the recovery variable $(y)$. When the bias is applied to the $x$ variable we take $s_{x} \neq 0$, $s_{y} = 0$ and when it is applied in the $y$ variable, we take $s_{x} = 0$, $s_{y} \neq 0$. $A_{ij}$ is the adjacency matrix and $A_{ij} = 1$ if $i \neq j$ and $A_{ij} = 0$ if $i=j$. In this network, parameters $a$ and $c$ are homogeneous whereas $b$ is considered to be heterogeneous as in Ref. \cite{r8}. 
Here, for our analysis we consider the following three systems as in Ref. \cite{r8}.
\begin{enumerate}[label=(\Alph*)]
	\item $Two-$coupled system, where $i = 1,2$, with parameters $a_{i} = -0.025794$, $b_{1} = 0.0065$, $b_{2} = 0.0135$, $c_{i} = 0.02$ and $k = 0.128$.
	\item $Three-$coupled system, where $i = 1,2,3$, with parameters $a_{i} = -0.0274546$, $c_{i} = 0.02$, $k = 0.064$ and the parameter $b$ is distributed from the expression $b_{i} = 0.006 + 0.008(\frac{i - 1}{n - 1})$. 
	\item $N-$coupled system, where $i = 1,2,...,101$, with parameters $a_{i} = -0.02651$, $c_{i} = 0.02$, $k = 0.00128$ and the parameter $b$ is distributed from the expression $b_{i} = 0.006 + 0.008(\frac{i - 1}{n - 1})$.
\end{enumerate}
\par System (2) with (A) and (B) configurations are numerically integrated using fourth order Runge-Kutta method (RK4) algorithm whereas with (C) configuration is numerically integrated using fourth-fifth order Runge-Kutta-Fehelberg method (RKF45) with adaptive step size.
\par Our aim is to mitigate extreme events. Before presenting the results, we will define what an extreme event is and the definition of various measures taken.
In this work, an event is extreme if time evolution of the system exceeds a threshold. The threshold is calculated using $x_{th}=  \langle \bar{x}_{m} \rangle + n\sigma_x$ \cite{r11}, where $\bar{x}_{m}$ is the first moment of the peaks, $\sigma_x=\sqrt{\big(\bar{x}^2)_{m}  - (\bar{x}_{m})^{2}}$ is the second moment over first moment and $4\leq n \leq 8$ \cite{r17}. The first and second moment refers to mean and the standard deviation. We fix $n=8$ throughout this work and estimate $x_{th}$. The value of threshold is calculated by collecting the peaks upto $10^{9}$. The mitigation of extreme events is verified from the $d_{max}$ and probability analysis. Here $d_{max}$ indicates the number of standard deviations that crosses the average amplitude corresponding to $\bar{x}_{max}$, where $\bar{x}_{max}$ indicates the global maximum. The $d_{max}$ value is calculated from the formula $d_{max}=\frac{max(\bar{x}) - \langle \bar{x} \rangle}{\sigma_x}$ \cite{r39}. The probability of extreme events is calculated as the ratio of number of peaks of $\bar{x}$ crossing the threshold to the total number of peaks of $\bar{x}$.
\section{Results}
In this section, we use constant bias and observe how extreme events are mitigated in $Two$, $Three$ and $N-$coupled FHN systems.
\subsection{Two-coupled system - mitigation of extreme events}  
\subsubsection{Negative constant bias in membrane potential $(x)$ variable} 
\begin{figure}[!ht]
	\centering
		\includegraphics[width=0.6\textwidth]{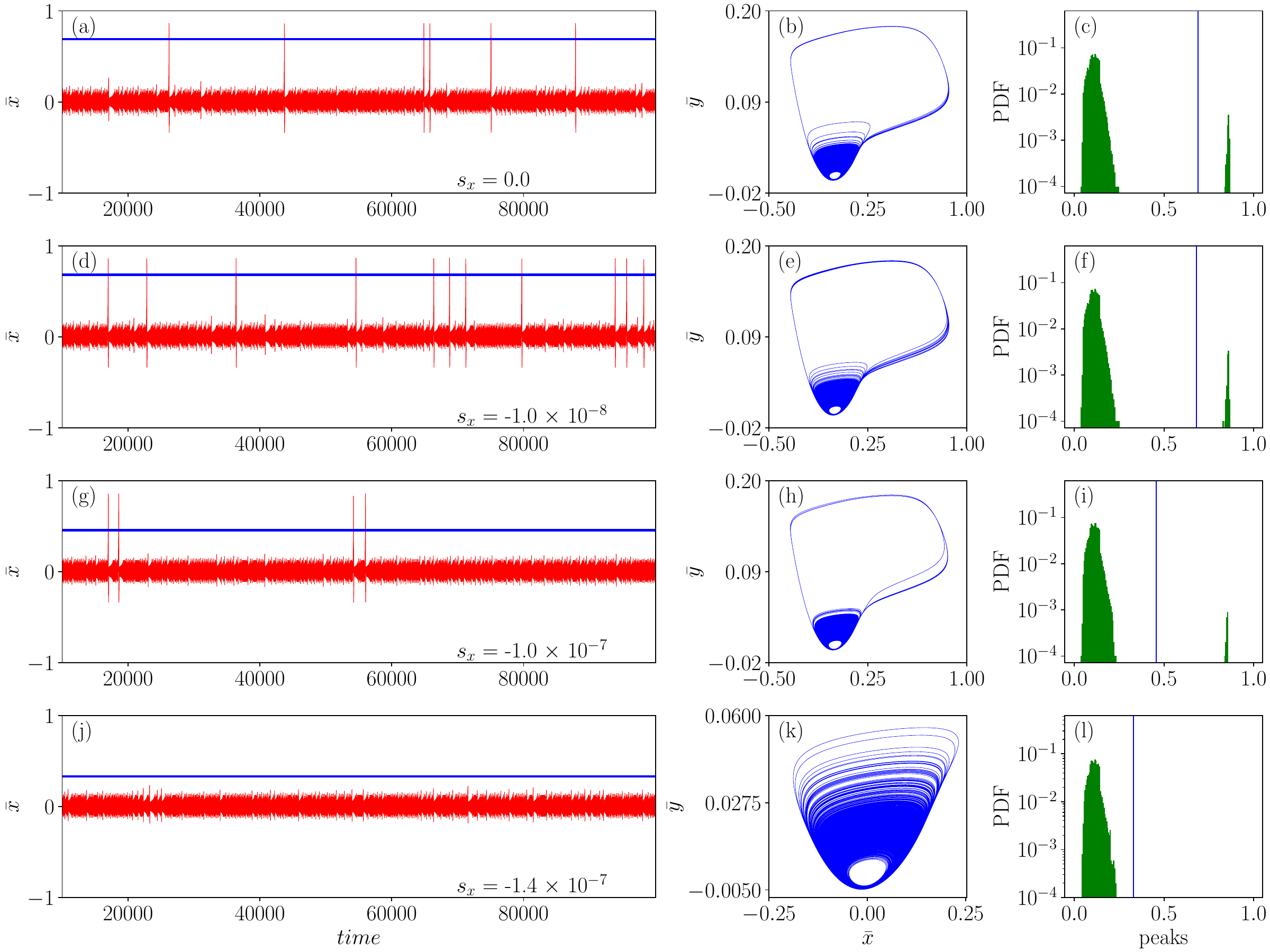}	
	\caption{The time evolution, phase portraits and probability distribution plots of system (2) is shown in left, middle and the right columns respectively. The parameters are $a_{1}=a_{2}=-0.025794$, $b_{1} = 0.0065$, $b_{2} = 0.0135$, $c_{1}=c_{2}=0.02$ and $k = 0.128$. From the first row to the fourth row, the values of constant bias are $s_{x}=0.0$, $s_{x} = -1.0 \times 10^{-8}$, $s_{x} = -1.0 \times 10^{-7}$ and $s_{x} = -1.4 \times 10^{-7}$ respectively. The blue line in the first column indicates the threshold.}
	\label{fhn_tspp_anx}
\end{figure}
\par Extreme events were observed in the system (2) for the set of parameters as mentioned earlier in (A) in Sec. 2. Figures 1(a) and 1(b) denote the time series and phase portrait of system (2) respectively for constant bias $s_{x}=0.0$. The large amplitude peaks crossing the threshold line (blue horizontal line in time series) are extreme events and the events can be visualized as a long excursion of trajectory from the bounded chaotic domain. Fig. 1(c) is the probability distribution function (PDF) of peaks and it confirms the presence of extreme events. For a slight increase in $s_{x} = -1.0 \times 10^{-8}$, we can see that the extreme events still sustains. This is verified from the plot in Fig. 1(d) as there are large amplitude peaks. There is also a long excursion of trajectory from the bounded chaotic domain in Fig. 1(e). We can notice from Fig. 1(f) that the probability for the extreme event to occur beyond the threshold has decreased. Now, when the bias value is increased to $s_{x} = -1.0 \times 10^{-7}$, the number of peaks crossing the threshold decreases appreciably. The decrease in large amplitude peaks can be observed from the plot in Fig. 1(g) and the long excursion is sustained which can be seen from Fig. 1(h). The probability of extreme events is decreased in the PDF plot as well, as observed from Fig. 1(i). Further increasing the bias value to $s_{x} = -1.4 \times 10^{-7}$, suppresses the extreme events entirely and only the low amplitude motion persists. The mitigation of extreme events can be verified from the Fig. 1(j) as there are no large amplitude peaks in time series. Further there is no long excursion of trajectory as confirmed from Fig. 1(k). We can confirm the mitigation of extreme events further from the PDF plot in the Fig. 1(l) as the probability for the event to occur beyond the threshold is zero even for very long time iterations. 
\begin{figure}[!ht]
	\centering
		\includegraphics[width=0.5\textwidth]{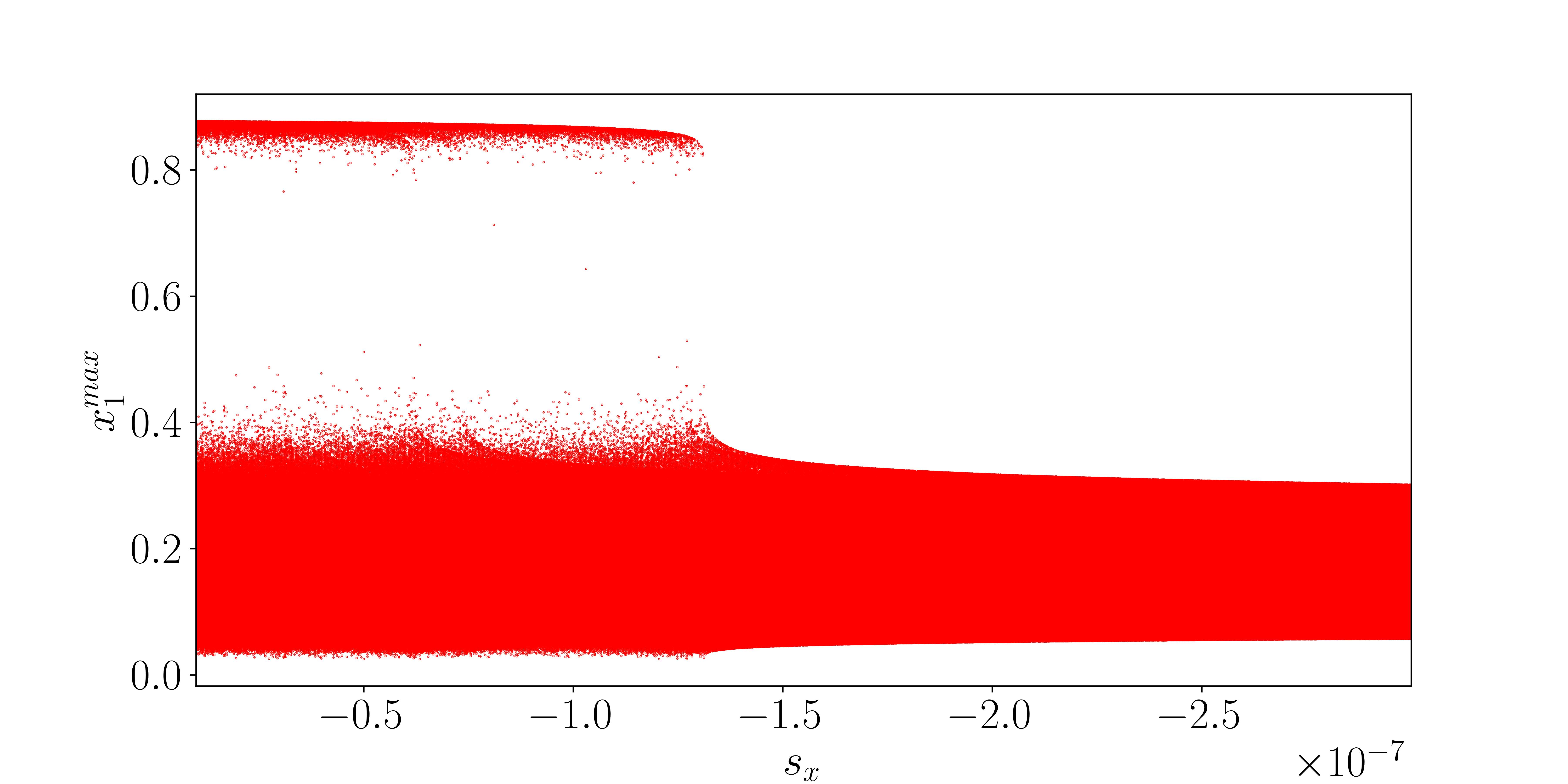}	
	\caption{Bifurcation plot of the first oscillator in system (2). Constant bias ($s_{x}$) is taken as the bifurcation parameter. The size of the chaotic attractor shrinks at $s_{x}=-1.4\times10^{-7}$. Other parameters are the same as used in Fig.~1.}
	\label{fhn_2c_bif_anx}
\end{figure}
\begin{figure}[!ht]
	\centering
	\includegraphics[width=0.5\textwidth]{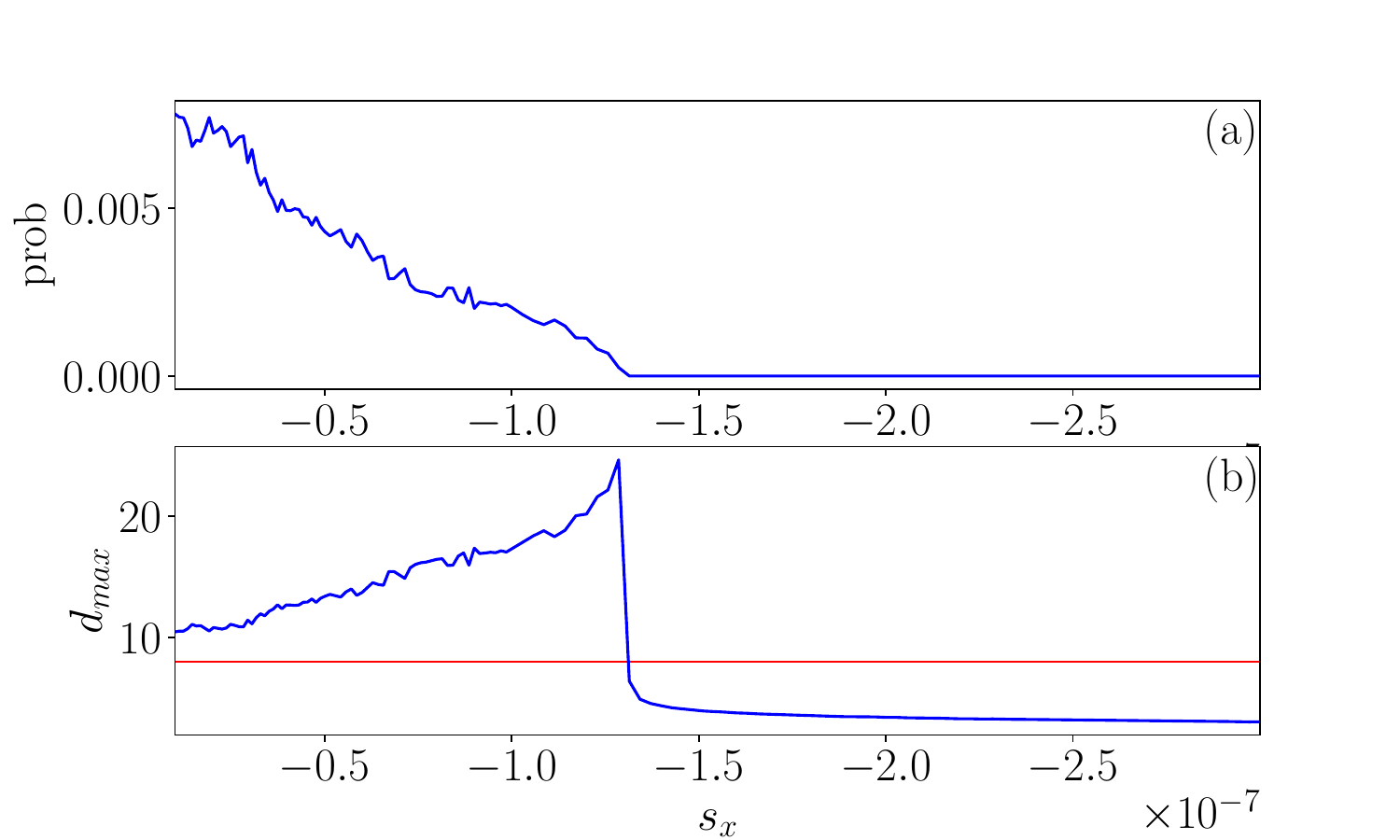}	
	\caption{(a) Variation of probability as a function of $s_{x}$  and (b) Variation of $d_{max}$ as a function of $s_{x}$ in system (2). The horizontal red line in (b) indicates the threshold $n=8$. The probability becomes zero and the $d_{max}$ is reduced below the red line at $s_{x}=-1.4\times10^{-7}$. Other parameters are the same as used in Fig.~1.}
	\label{fhn_2c_dmpb_anx}
\end{figure}
\par The local maxima of $x_{1}$ as a function of $s_{x}$  is shown in Fig. 2. This is plotted by collecting the local maxima of $x_{1}$ of system (2) by varying constant bias $s_{x}$. The constant bias has no impact on extreme events upto $s_{x} = -1.3\times10^{-7}$. Suppression of extreme events occurs at $s_{x} = -1.4\times10^{-7}$ through decrease in size of the chaotic attractor. After this, the chaotic attractor transits into periodic attractor through three major bifurcations, namely (i) inverse tangent (saddle-node) bifurcation, (ii) inverse band-merging bifurcation and (iii) inverse period-doubling bifurcation \cite{r40,r41}. For clear visualization, we have shown only the transition of suppression of extreme events to chaos in all the cases. Finally, for further increase in constant bias, the chaotic dynamics transits into periodic dynamics  through reverse period-doubling bifurcation. 
\par The probability and $d_{max}$ analysis were done to confirm the mitigation of extreme events. The existence of probability of extreme events decreases as we increase the bias value which can be seen from Fig. 3(a). This occurs until $s_{x}=-1.3 \times 10^{-7}$. A slight increase in $s_{x}=-1.4 \times 10^{-7}$ leads to the complete suppression of extreme events. From Fig. 3(b), we can see that whenever the $d_{max}$ crosses the value $n = 8$, there is a probability for existence of extreme events. Increasing the bias value reduces the $d_{max}$ and the probability becomes zero. 
\subsubsection{Positive constant bias in recovery $(y)$ variable}
In this sub-section we are implementing positive constant bias in the recovery $y$ variable to mitigate extreme events.
\begin{figure}[!ht]
	\centering
		\includegraphics[width=0.6\textwidth]{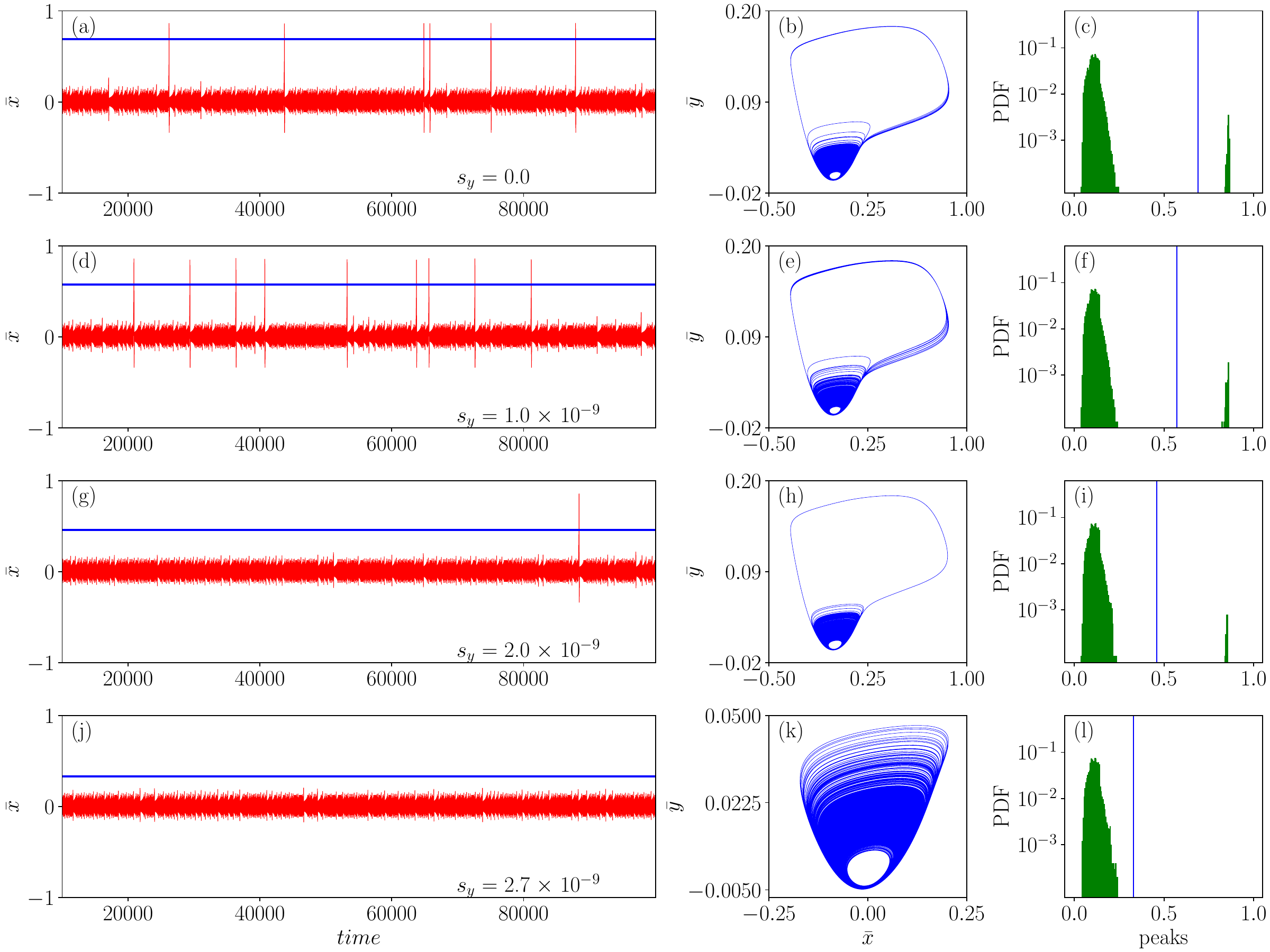}
	\caption{The left, middle and the right columns shows the time evolution, phase portraits and probability distribution plots of system (2). The parameters are $a_{1}=a_{2}=-0.025794$, $b_{1} = 0.0065$, $b_{2} = 0.0135$, $c_{1}=c_{2}=0.02$ and $k = 0.128$. From the first row to the fourth row, the value of constant bias are $s_{y}=0.0$, $s_{y} = 1.0 \times 10^{-9}$, $s_{y} = 2.0 \times 10^{-9}$ and $s_{y} = 2.7 \times 10^{-9}$ respectively. The blue line in the first column indicates the threshold.}
	\label{fhn_2c_tspp_apy}
\end{figure}
\par The parameters are same as in the case of negative bias. The first row in Fig. 4 corresponds to time series, phase portrait and PDF plots of system (2) for a constant bias, $s_{y} = 0.0$. The large amplitude peaks crossing the threshold are extreme events and are observed from Fig. 4(a). The long excursion of the trajectory in Fig. 4(b) corresponds to the extreme event. The existence of extreme events is confirmed from the peak PDF plotted in Fig. 4(c). Next, for the value of constant bias $s_{y} = 1.0 \times 10^{-9}$, we can notice that the extreme events sustains. The time series plot in Fig. 4(d) shows that there are large amplitude peaks and a long excursion of trajectory from the bounded chaotic domain in Fig. 4(e). The probability beyond the threshold is reduced as we can see from Fig. 4(f). A slight increase in bias, $s_{y} = 2.0 \times 10^{-9}$ considerably decreases the number of peaks crossing the threshold and is observed from the plot in Fig. 4(g). The corresponding decrease in the long excursion of the trajectory from the bounded chaotic domain is shown in Fig. 4(h). Reduction in probability of events beyond the threshold is shown in Fig. 4(i). Then increasing the bias value to $s_{y} = 2.7 \times 10^{-9}$, suppresses the extreme events entirely and only the bounded chaotic motion persists as similar to the previous case. The complete mitigation of extreme events can be seen from Fig. 4(j) as no large amplitude peaks are present in the time series. Correspondingly in Fig. 4(k), there is no long excursion in the trajectory and no probabilities of occurence of extreme events beyond the threshold.
\begin{figure}[!ht]
	\centering
		\includegraphics[width=0.5\textwidth]{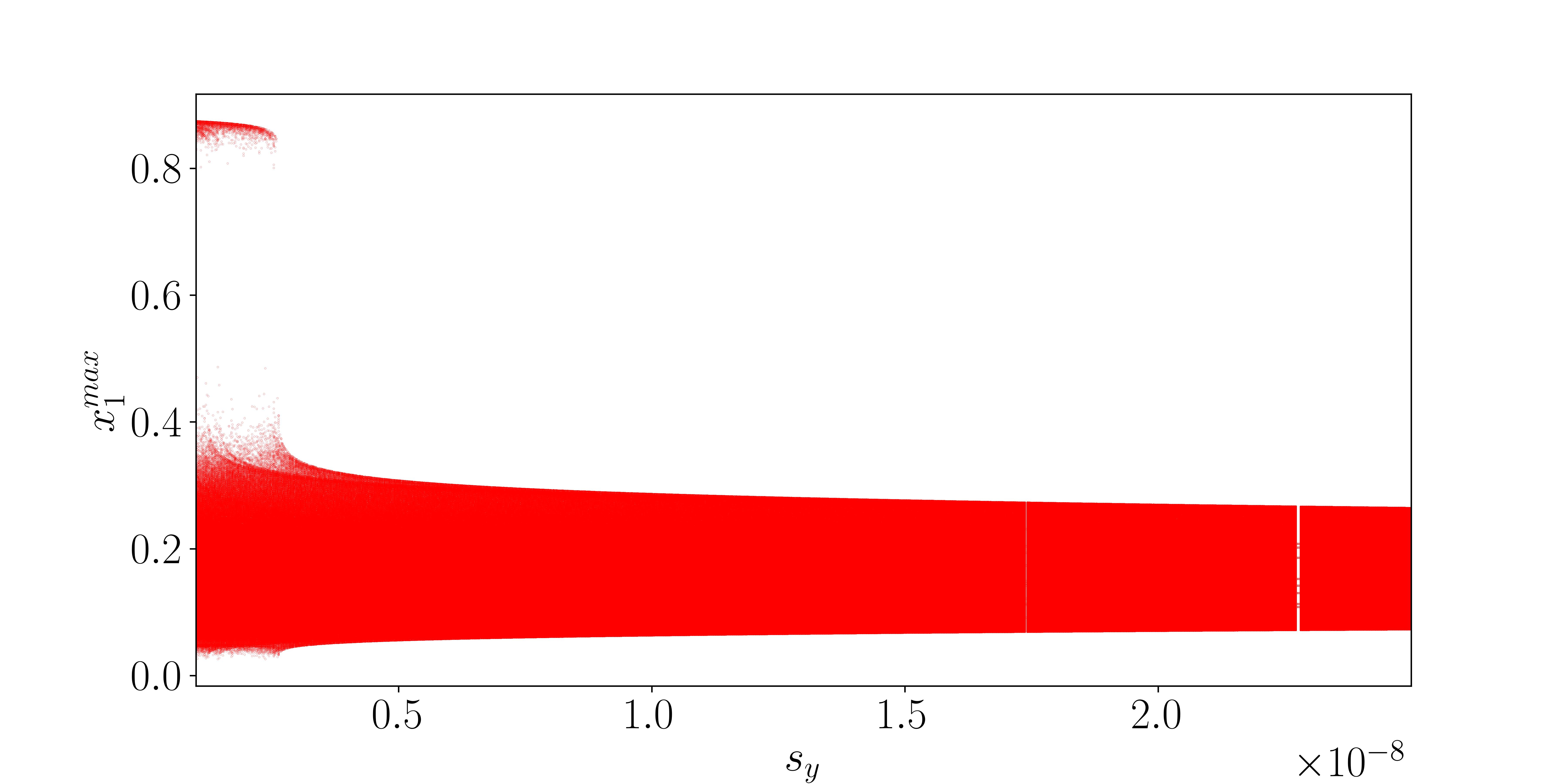} 
	\caption{Local maxima of oscillator-1 in system (2) with respect to constant bias ($s_{y}$). The shrinkage of chaotic attractor occurs at $s_{y}=0.27\times10^{-8}$. Other parameters are the same as used in Fig.~4.}
	\label{fhn_2c_bif_apy}
\end{figure}
\par In Fig. 5, we have depicted the bifurcation diagram of system (2) with varying $s_{y}$. From the figure, we can observe that upon introducing constant bias and then increasing its value decreases the size of the chaotic attractor. There is no decrease in size of the chaotic attractor upto $s_{y} = 0.26\times10^{-8}$. Then increasing bias to $s_{y} = 0.27\times10^{-8}$ suddenly decreases the size of the chaotic attractor which leads to the suppression of extreme events and only chaotic dynamics remains. There occurs a periodic window at $s_{y} \approx 2.3\times10^{-8}$ in between the chaotic regime through the inverse tangent bifrucation.
\begin{figure}[!ht]
	\centering
		\includegraphics[width=0.5\textwidth]{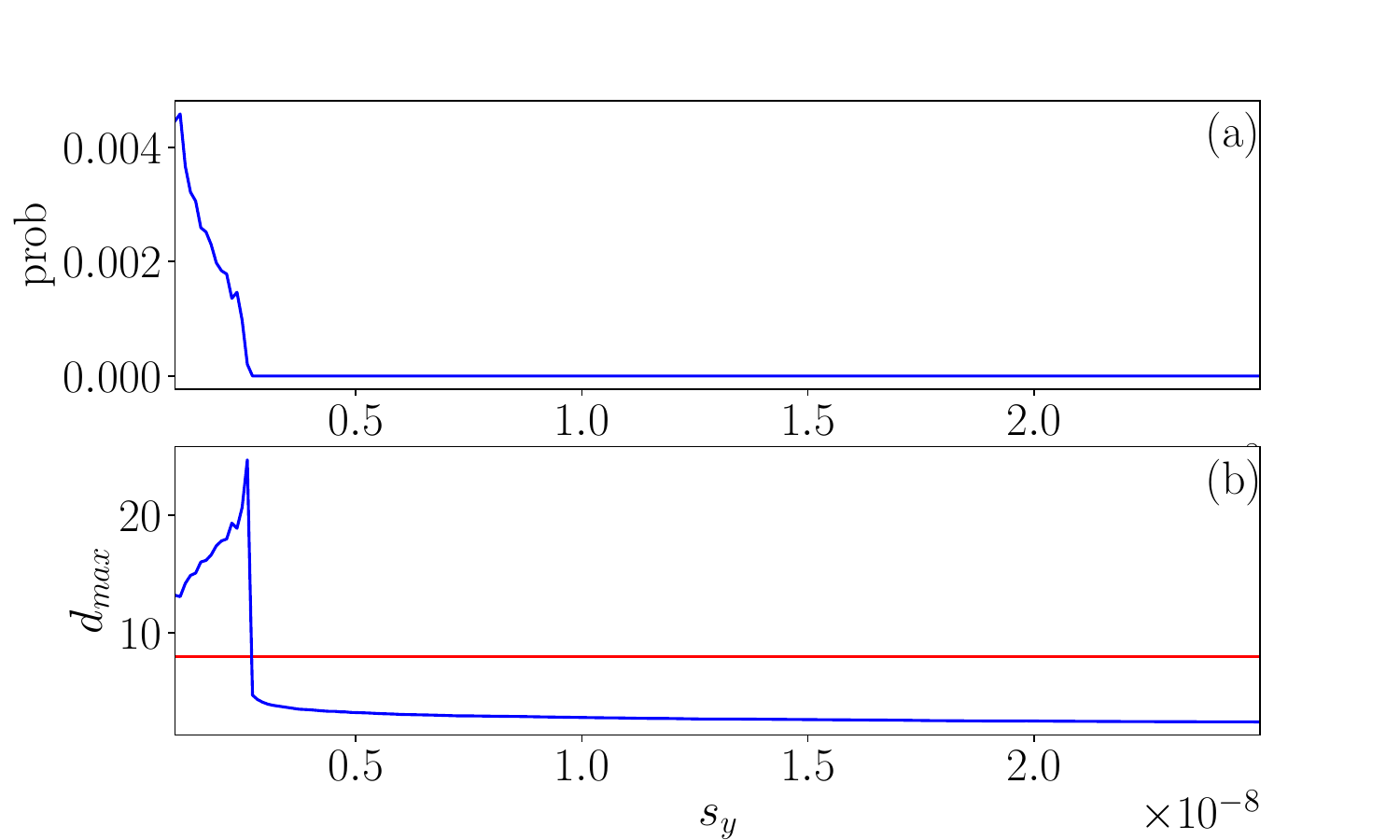}	
	\caption{(a) Probability plot and (b) $d_{max}$ plot with respect to $s_{y}$ in system (2). For the value of constant bias $s_{y}=0.27\times10^{-8}$, the $d_{max}$ decreases below $n=8$ line (red line) and the probability becomes zero. Other parameters are the same as used in Fig.~4.}
	\label{fhn_2c_dmpb_apy}
\end{figure}
\par The suppression of extreme events by varying $s_{y}$ is further confirmed from the probability and $d_{max}$ plots. When the value of bias is  $s_{y}=0.27 \times 10^{-8}$, the probability becomes zero and the $d_{max}$ is reduced below the $n=8$ line which can be seen from Figs. 6(a) and 6(b) respectively. Increasing the bias value decreases $d_{max}$ and the probability also becomes zero.
\subsection{Three-coupled system - emergence of extreme events}
\par Next, we consider the $Three-$coupled system using equation (2) with the set of parameters given in (B). Before explaining the mitigation of extreme events, we need to assure the existence of extreme events for the three coupled case. Hence, we considered the parameters $a$ and $c$ to be homogeneous and $b$ to be heterogenous as in Sec. 3.1. It has been found that the three coupled FHN model exhibits extreme events at $k=0.064$ which can be observed from the Fig. 7(a). The blue line in plot of Fig. 7(a) is the threshold $x_{th}$ and the large amplitude peaks crossing the threshold are extreme events. Figure 7(b) represents the phase portrait and we can observe that the long expedition of the orbit from the bounded chaotic motion is the extreme event. The emergence of extreme event is verified from the PDF plot in Fig. 7(c) where there is a finite non-zero probability after the threshold. 
\begin{figure}[!ht]
	\centering
	\includegraphics[width=0.7\textwidth]{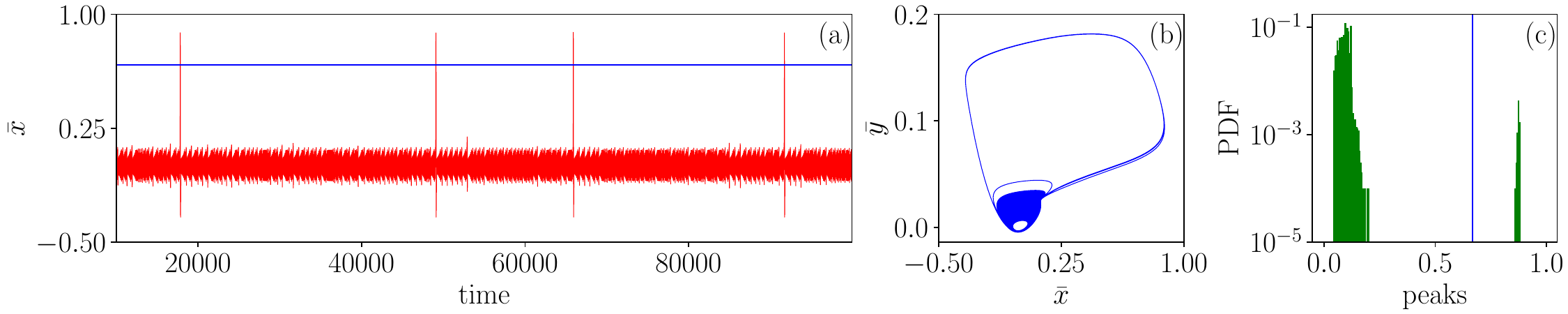}	
	\caption{(a) Time series in $x$, (b) phase portraits and (c) probability distribution plots of system (2) for three coupled case. The parameters are $a_{1}=a_{2}=a_{3}=-0.0274546$, $c_{1}=c_{2}=c_{3}=0.02$, $k = 0.064$. The parameter $b$ is distributed from the expression $b_{i} = 0.006 + 0.008(\frac{i - 1}{n - 1})$. The horizontal blue line in (a) indicates the threshold.}
	\label{fhn_3c_emer}
\end{figure}
\begin{figure}[!ht]
	\centering
		\includegraphics[width=0.5\textwidth]{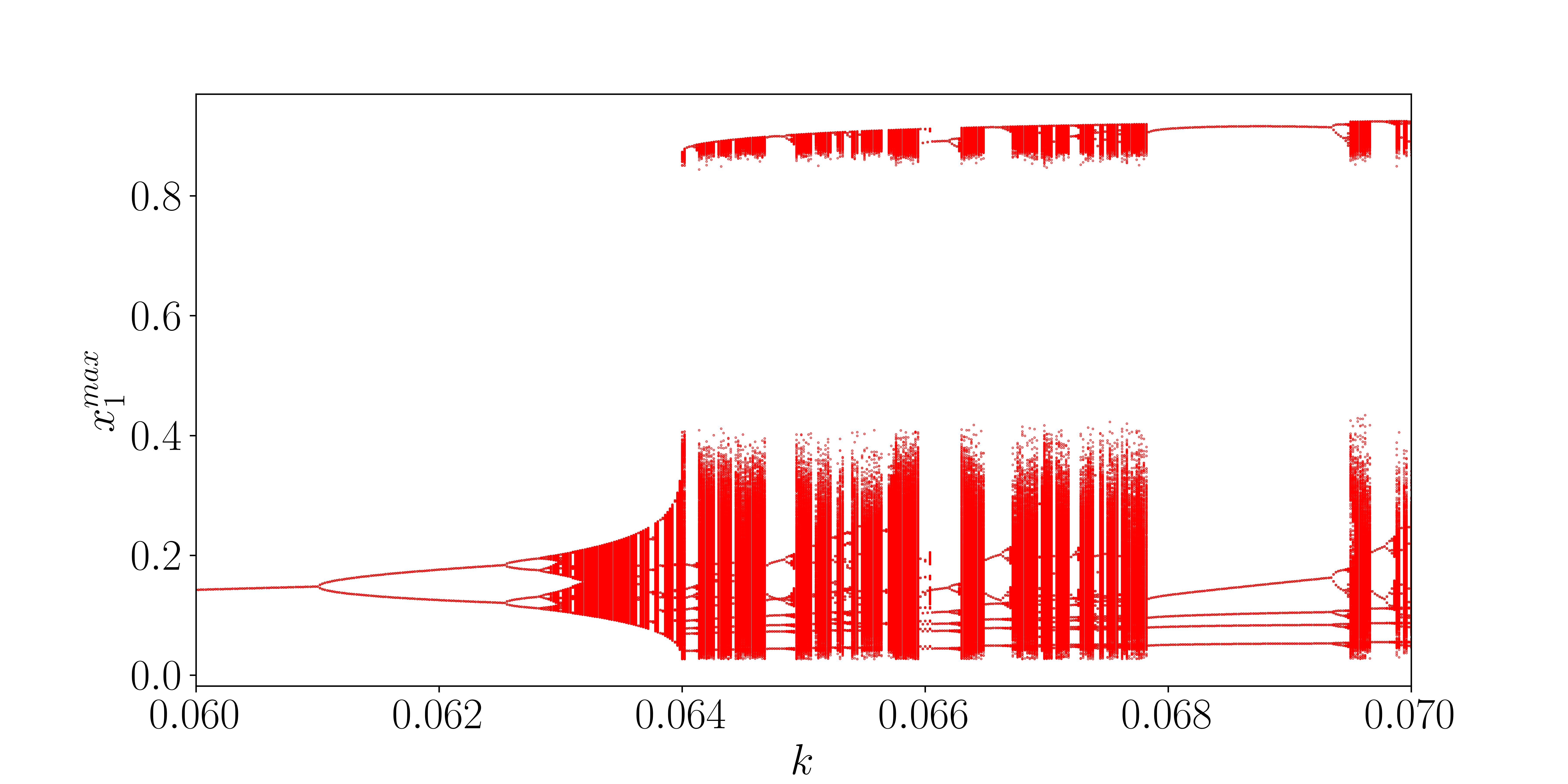}
	\caption{Peaks of the first oscillator in system (2) with parameter set (B) plotted with respect to the coupling strength ($k$). The interior crisis in the system occurs at $k=0.064.$ Other parameters are the same as used in Fig.~7.}
	\label{fhn_3c_bif}
\end{figure}
\begin{figure}[!ht]
	\centering
	\includegraphics[width=0.5\textwidth]{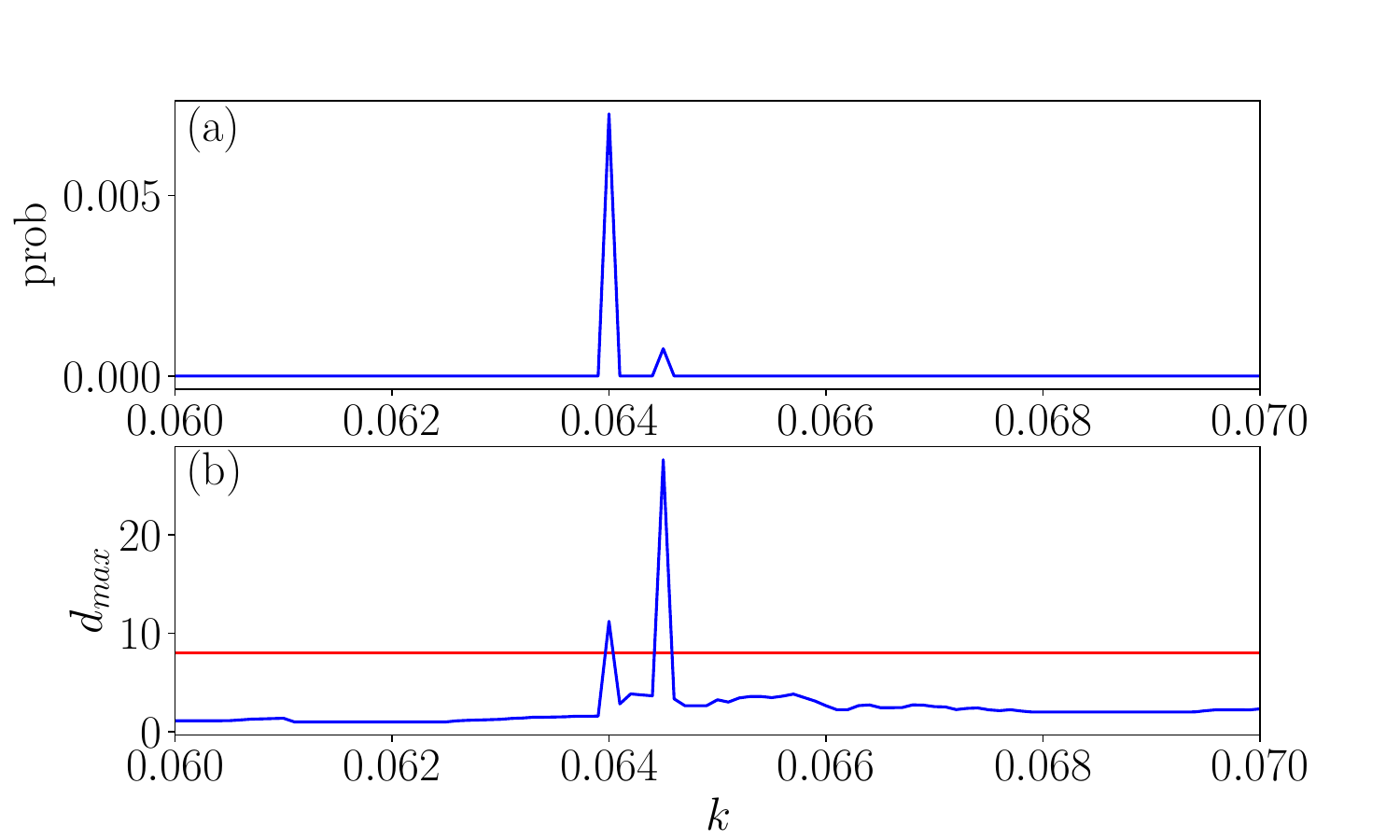}
	\caption{(a) Probability and (b) $d_{max}$ plot varying as a function of $k$ in system (2). The horizontal red line in (b) indicates the $n=8$ line. The probability increases and the $d_{max}$ crosses the threshold at $k=0.064$ and $k=0.0645$ which are the points of emergence of extreme events. Other parameters are the same as used in Fig.~7.}
	\label{fhn_3c_dmpb}
\end{figure}
\par In Fig. 8, we plot the bifurcation diagram of system (2) with $n = 3$. From Fig. 8, we can observe that for the three coupled FHN system the route to chaos occurs through period-doubling bifurcation and band merging bifurcation. At $k\approx0.06318$ the chaotic attractor consists of two bands. These two bands merge together at $k=0.0632$ and forms a single band chaotic attractor. At $k=0.064$, interior crisis occurs in the system and there is a sudden expansion of the bounded chaotic attractor. At this crisis point, sudden large amplitude oscillations occur in the system which leads to the generation of extreme events. Due to the opening and shutting of the channel-like structure in the state space \cite{r15}, the inaccessible parts of the state space before the crisis point have now become accessible for the trajectories after the crisis point. Hence the trajectories take a long excursion in the state space which can be observed from the plot in Fig. 7(b). Following this, through saddle-node bifurcation, periodic windows occur in the chaotic regime. These periodic oscillations comprise of both the small and large amplitude oscillations known as mixed-mode oscillations. Similar to Ref. [15] mixed-mode oscillations also occur in the chaotic regime after the crisis point in the three coupled system. Apart from the crisis point $k=0.064$, extreme events were observed in the chaotic regime only at $k = 0.0645$. These observations are illustrated in Fig. 8. 
\par To confirm our observation of emanation of extreme events, probability and $d_{max}$ analysis are also carried out. From Fig. 9(a), at $k=0.064$ and $k=0.0645$ there is a probability of existence of extreme events. Hence, whenever there is a probability of existence of extreme events, the $d_{max}$ crosses the $n$ line and at other places the probability is zero and $d_{max}$ is lower than the $n$ line.  
\subsection{Mitigation of extreme events}
\par Having seen the emanation of extreme events in the $Three-$coupled case, we intend to mitigate it as in the $Two-$coupled case. The parameters are the same as in the previous sub-section.
\subsubsection{Negative constant bias in membrane potential $(x)$ variable}
\begin{figure}[!ht]
	\centering
		\includegraphics[width=0.6\textwidth]{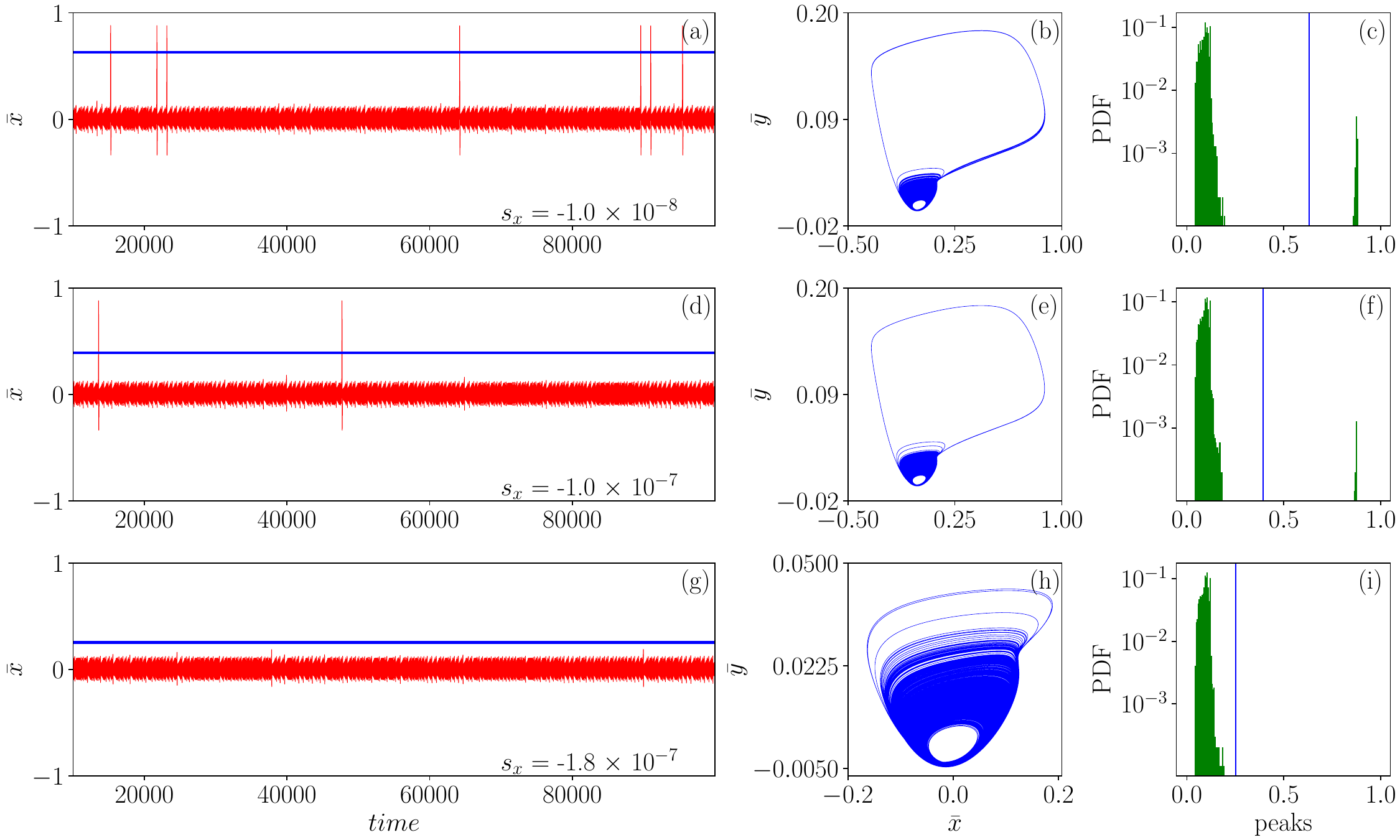}
	\caption{Time series in $x$ (column 1), phase portraits (column 2) and PDF plot (column 3) of system (2). The parameters are $a_{1}=a_{2}=a_{3}=-0.0274546$, $c_{1}=c_{2}=c_{3}=0.02$, $k = 0.064$. Parameter $b$ is distributed from the expression $b_{i} = 0.006 + 0.008(\frac{i - 1}{n - 1})$. From the first row to the third row, the value of constant bias are $s_{x} = -1.0 \times 10^{-8}$, $s_{x} = -1.0 \times 10^{-7}$ and $s_{x} = -1.8 \times 10^{-7}$ respectively. The blue line in the first column indicates the threshold.}
	\label{fhn_3c_tspp_anx}
\end{figure}
\par For the value of constant bias, $s_{x} = -1.0 \times 10^{-8}$, the extreme events sustains in system (2) and is observed from the time series plot in (Fig. 10(a)) where the large amplitude peaks crossing the threshold line are extreme events and correspondingly long excursion of trajectory from chaotic domain corresponds to the extreme event in Fig. 10(b). The extreme events can also be confirmed from the PDF plot in Fig. 10(c) as there is probability after the threshold. Then the bias value is increased to $s_{x} = -1.0 \times 10^{-7}$ and we find that the number of large amplitude peaks is decreased. The decrement in peaks crossing the threshold and the long excursion of the trajectory can be seen from the plots in Figs. 10(d) and 10(e). The decrease in probability is observed from the PDF plot in Fig. 10(f). Finally, increasing the bias value to $s_{x} = -1.8 \times 10^{-7}$, mitigates the extreme events and only the chaotic motion persists. The mitigation of extreme events can be observed from the time series in Fig. 10(g), phase portrait in Fig. 10(j) and PDF plot in the Fig. 10(i) because there is no probability beyond the threshold.
\begin{figure}[!ht]
	\centering
		\includegraphics[width=0.5\textwidth]{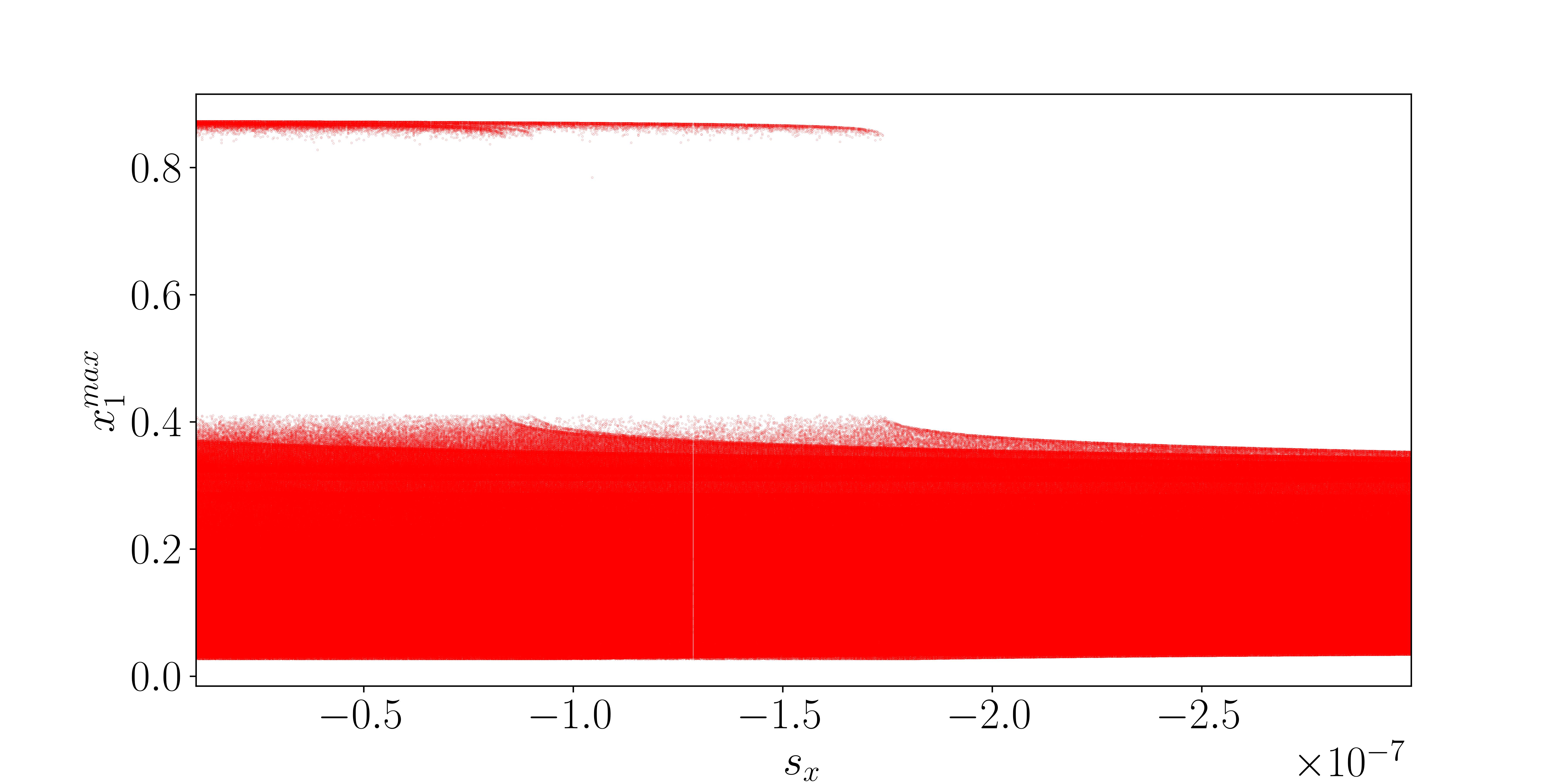}
	\caption{Bifurcation diagram drawn by collecting the peaks of the first oscillator in system (2) as a function of constant bias ($s_{x}$). The size of the chaotic attractor gets reduced at $s_{x}=-1.8\times10^{-7}$. Other parameters are the same as used in Fig.~10.}
	\label{fhn_3c_bif_anx}
\end{figure}
\begin{figure}[!ht]
	\centering
		\includegraphics[width=0.5\textwidth]{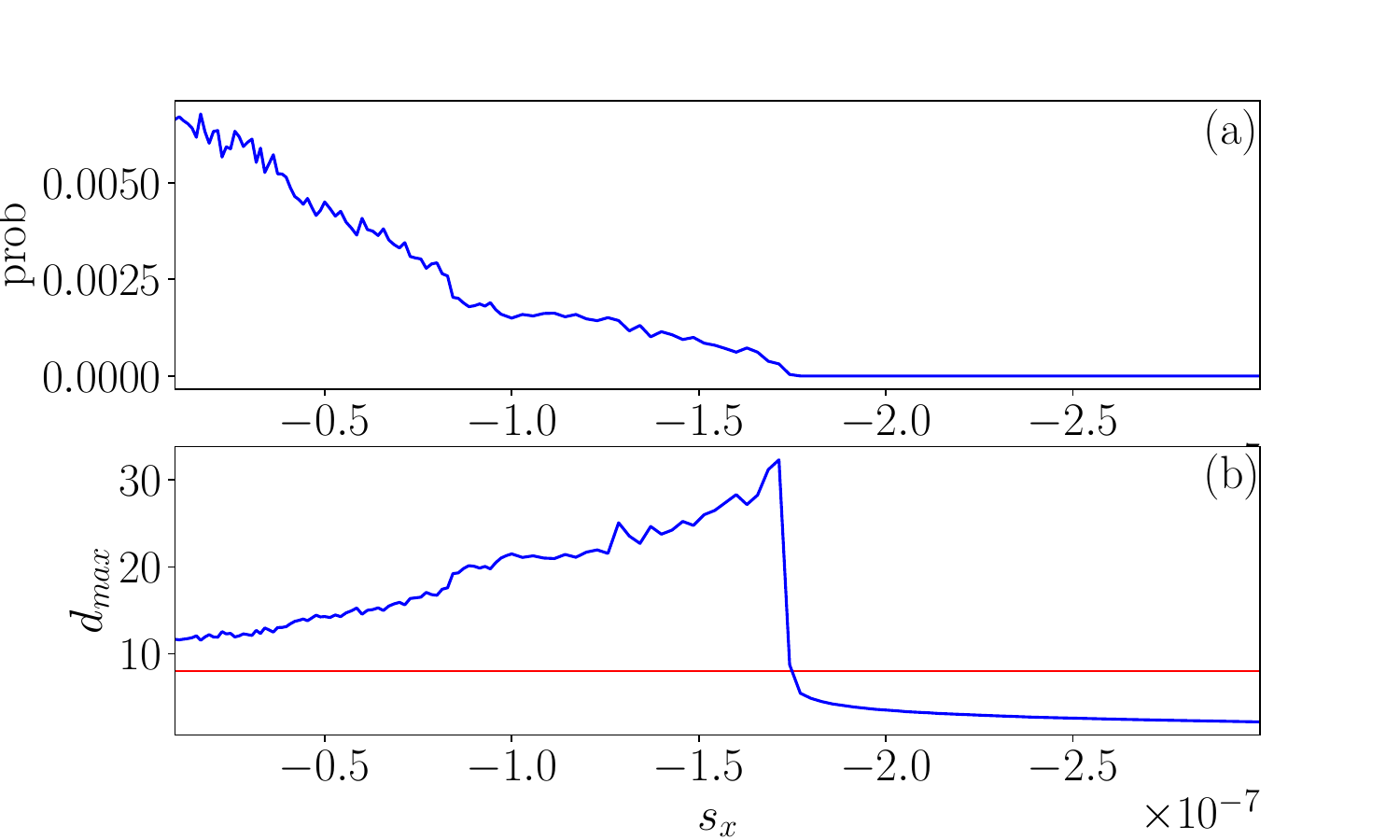}
	\caption{(a) Probability plot and (b) $d_{max}$ plot for varying $s_{x}$ in system (2). The probability becomes zero and the $d_{max}$ is reduced below $n=8$ line (red line) at $s_{x}=-1.8\times10^{-7}$. Other parameters are the same as used in Fig.~10.}
	\label{fhn_3c_dmpb_anx}
\end{figure}
\par In Fig. 11, we plot the bifurcation diagram of the first oscillator in system (2) for varying $s_{x}$. The size of the chaotic attractor decreases as we increase the bias in the negative spatial direction. Extreme events sustain and the constant bias has no impact upto $s_{x} = -1.7\times10^{-7}$. Suppression of extreme events occurs at $s_{x} = -1.8\times10^{-7}$ through sudden decrease in size of the chaotic attractor. 
\par The mitigation of extreme events for varying $s$ is further confirmed from the probability and $d_{max}$ analysis. There is no change in the existence of extreme events until $s_{x}=-1.7 \times 10^{-7}$. The $d_{max}$ value decreases below the threshold line when the bias value is slightly increased to $s_{x}=-1.8 \times 10^{-7}$ which can be seen from the plot in Fig. 12(b). Also the probability becomes zero (Fig. 12(a)). Further increasing the bias value reduces the $d_{max}$ and one can notice that the probability becomes zero. 
\subsubsection{Positive constant bias in recovery $(y)$ variable}
\par System (2) exhibits an extreme event for the bias value $s_{y} = 1.0 \times 10^{-10}$. A slight increase in bias $s_{y} = 1.0 \times 10^{-9}$ does not affect the extreme events and the number of large amplitude peaks remains the same (Figs. 13(a) and 13(d)). Extreme event is observed as the long expedition of the orbit from the bounded chaotic domain in Figs. 13(b) and 13(e). The extreme events can be confirmed from the PDF plot in Fig. 13(c) and 13(f) as there is a probability after the threshold. Finally increasing the bias value to $s_{y} = 4.0 \times 10^{-9}$ mitigates the extreme events and only the chaotic motion persists (Figs. 13(g) and 13(h)). The mitigation of extreme events is verified from the PDF plot given in the Figs. 13(i). 
\par Next, we plotted the bifurcation of the first oscillator in system (2) with the parameters set (B) for varying $s_y$ in Fig. 14. Increasing the constant bias value decreases the size of the chaotic attractor. There is no decrease in size of the chaotic attractor upto $s_{y} = 0.39\times10^{-8}$. However, increasing bias to $s_{y} = 0.4\times10^{-8}$ decreases the size of the chaotic attractor which leads to the suppression of extreme events and only chaotic dynamics sustains. Subsequently, chaotic dynamics transits to periodic dynamics. Inverse tangent bifurcation is responsible for the generation of periodic dynamics in between the chaotic regimes at $s_{y} \approx 0.6\times10^{-8}$ and $ 1.25\times10^{-8}$. 
\par Probability and $d_{max}$ analysis were done to verify the mitigation of extreme events. The probability of existence of extreme events reduced when the bias is increased to $s_{y}= 0.4 \times 10^{-8}$ (Fig. 15(a)). Whenever there is a non-zero probability of existence of extreme events, the $d_{max}$ crosses the $n$ value of the threshold line (Fig. 15(b)). Increasing the bias value reduces $d_{max}$ and the probability becomes zero.
\begin{figure}[!ht]
	\centering
	\includegraphics[width=0.6\textwidth]{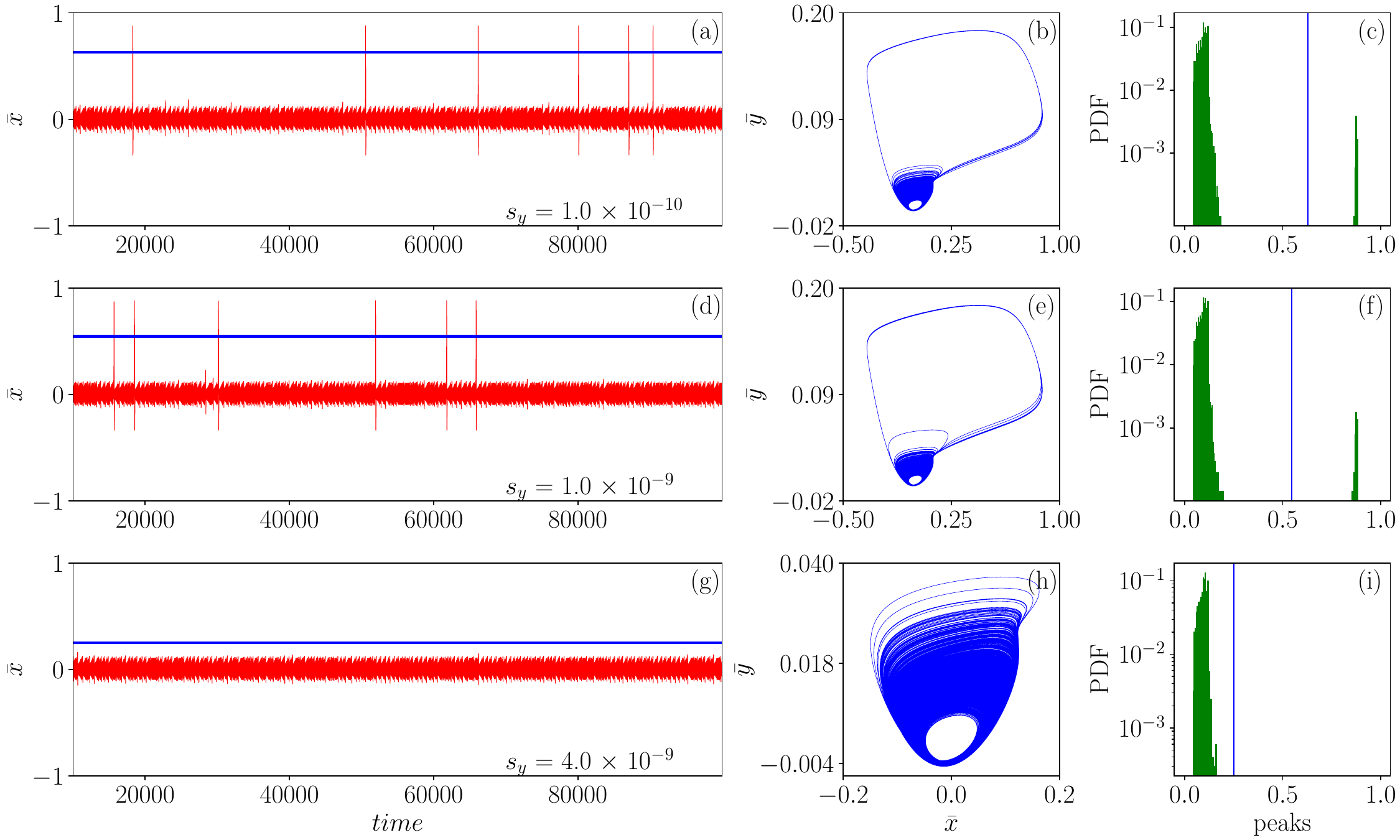}
	\caption{Time series in $x$, phase portraits in ($x-y$) plane and PDF of peaks system (2) are shown in left, middle and right columns respectively. The parameters are $a_{1}=a_{2}=a_{3}=-0.0274546$, $c_{1}=c_{2}=c_{3}=0.02$, $k = 0.064$ and the parameter $b$ is distributed from the expression $b_{i} = 0.006 + 0.008(\frac{i - 1}{n - 1})$. From the first row to the third row, the values of constant bias are $s_{y} = 1.0 \times 10^{-10}$, $s_{y} = 1.0 \times 10^{-9}$ and $s_{y} = 4.0 \times 10^{-9}$ respectively. The blue line in the first column indicates the threshold.}
	\label{fhn_3c_tspp_apy}
\end{figure}
\begin{figure}[!ht]
	\centering
		\includegraphics[width=0.5\textwidth]{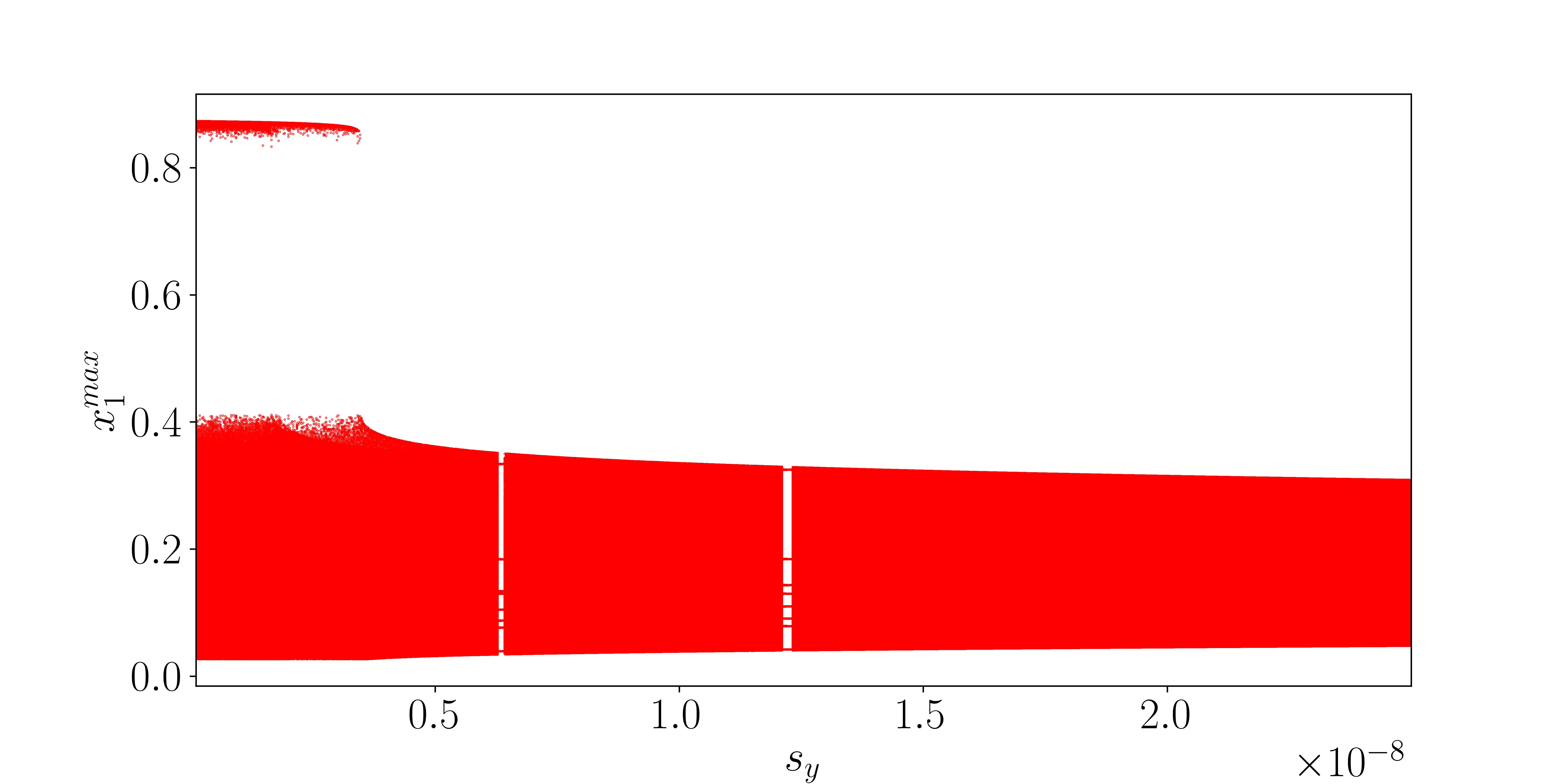}
	\caption{Local maxima of the first oscillator in system (2) with respect to constant bias ($s_{y}$). The reduction in the size of chaotic attractor occurs at $s_{y}=0.4\times10^{-8}$. Other parameters are the same as used in Fig.~13.}
	\label{fhn_3c_bif_apy}
\end{figure}
\begin{figure}[!ht]
	\centering
		\includegraphics[width=0.5\textwidth]{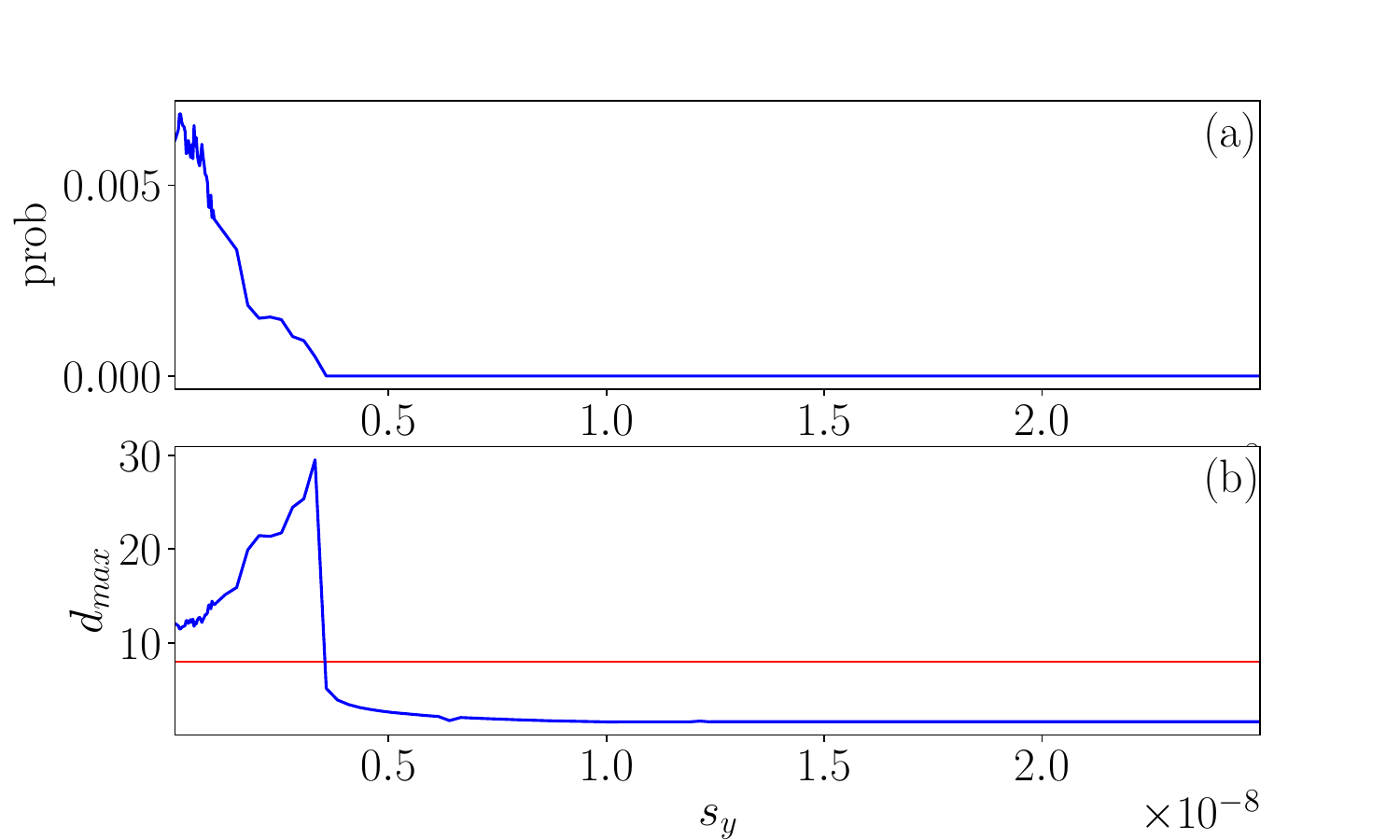}
	\caption{(a) Probability and (b) $d_{max}$ plot as a function of $s_{y}$ in system (2). The $d_{max}$ gets reduced below $n=8$ line and the probability becomes zero at $s_{y}=0.4\times10^{-8}$. Other parameters are the same as used in Fig.~13.}
	\label{fhn_3c_dmpb_apy}
\end{figure}
\subsection{N-coupled system - mitigation of extreme events}
Next, we extend the same procedure to mitigate extreme events in $N-$coupled FHN model and determine the route of suppression. 
\subsubsection{Negative constant bias in membrane potential $(x)$ variable}
\begin{figure}[!ht]
	\centering
		\includegraphics[width=0.6\textwidth]{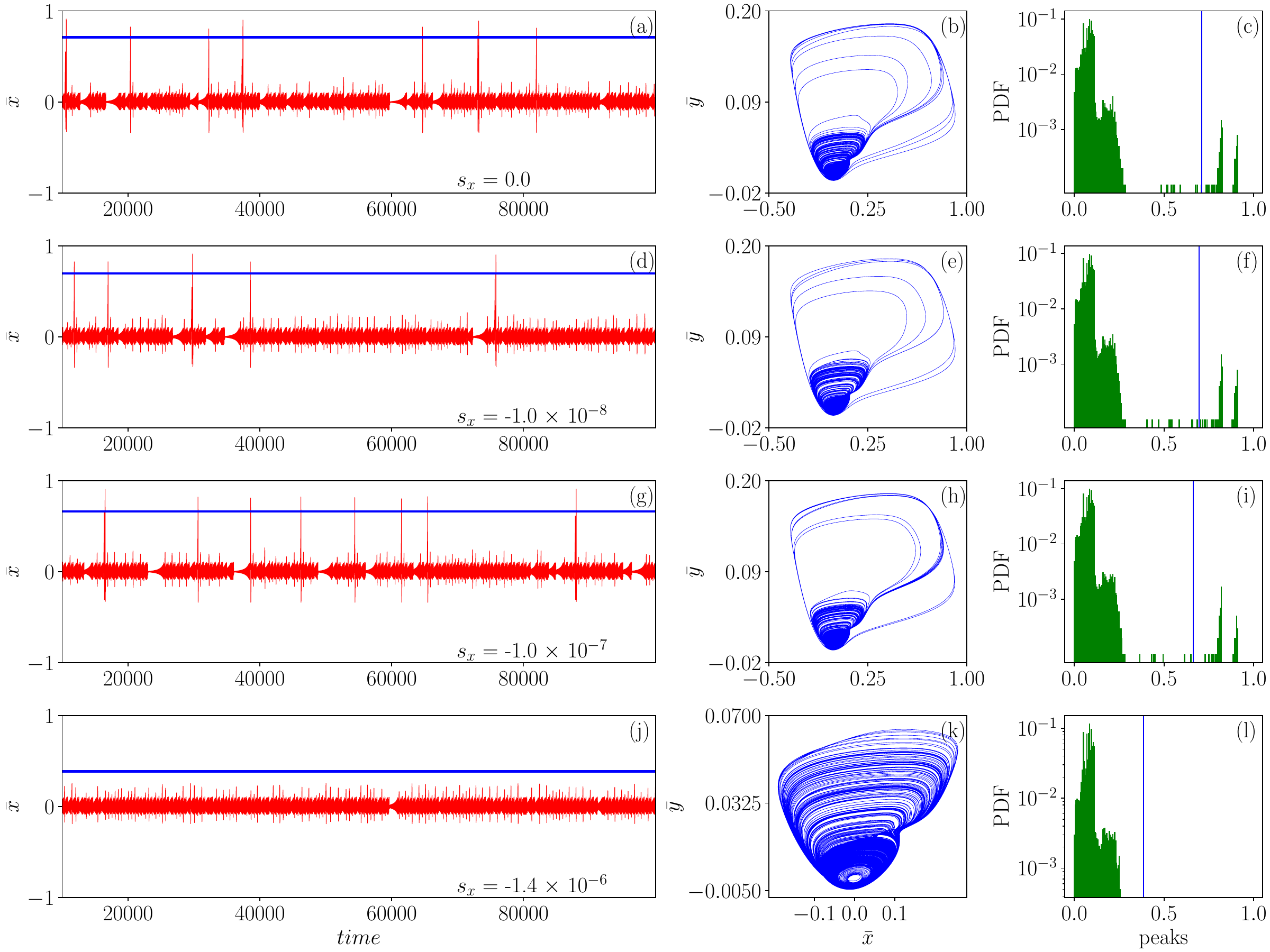}
	\caption{The time evolution (left column), phase portraits (middle column) and probability distribution plots (right column) of system (2). The parameters are $a_{i}=-0.02651$, $c_{i}=0.02$, $k = 0.00128$. Parameter $b$ is distributed from the expression $b_{i} = 0.006 + 0.008(\frac{i - 1}{n - 1})$. From the first row to the fourth row, the values of constant bias are $s_{x}=0.0$, $s_{x} = -1.0 \times 10^{-8}$, $s_{x} = -1.0 \times 10^{-7}$ and $s_{x} = -1.4 \times 10^{-6}$ respectively. The blue line in the first column indicates the threshold.}
	\label{fhn_nc_tspp_anx}
\end{figure}
\par Extreme events were observed in the system (2) for (C) set of parameters as mentioned earlier. The time series and the phase portrait of system (2) for constant bias $s_{x}=0.0$ is demonstrated in Figs. 16(a) and 16(b). The large amplitude peaks crossing the threshold line and the long excursion of trajectory from the bounded chaotic regime corresponds to the extreme event. These events are also confirmed from the PDF plot in Fig. 16(c). For the value of constant bias, $s_{x} = -1.0 \times 10^{-8}$ the extreme events still persist in the system. Now when the bias value is increased to $s_{x} = -1.0 \times 10^{-7}$, extreme events sustain but the number of large amplitude peaks reduced. Finally, increasing the bias value to $s_{x} = -1.4 \times 10^{-6}$ the extreme events get suppressed and only the chaotic motion persists. Our investigations reveal that in N-coupled system the extreme events get suppressed in the same way as in two and three coupled system.
\begin{figure}[!ht]
	\centering
		\includegraphics[width=0.5\textwidth]{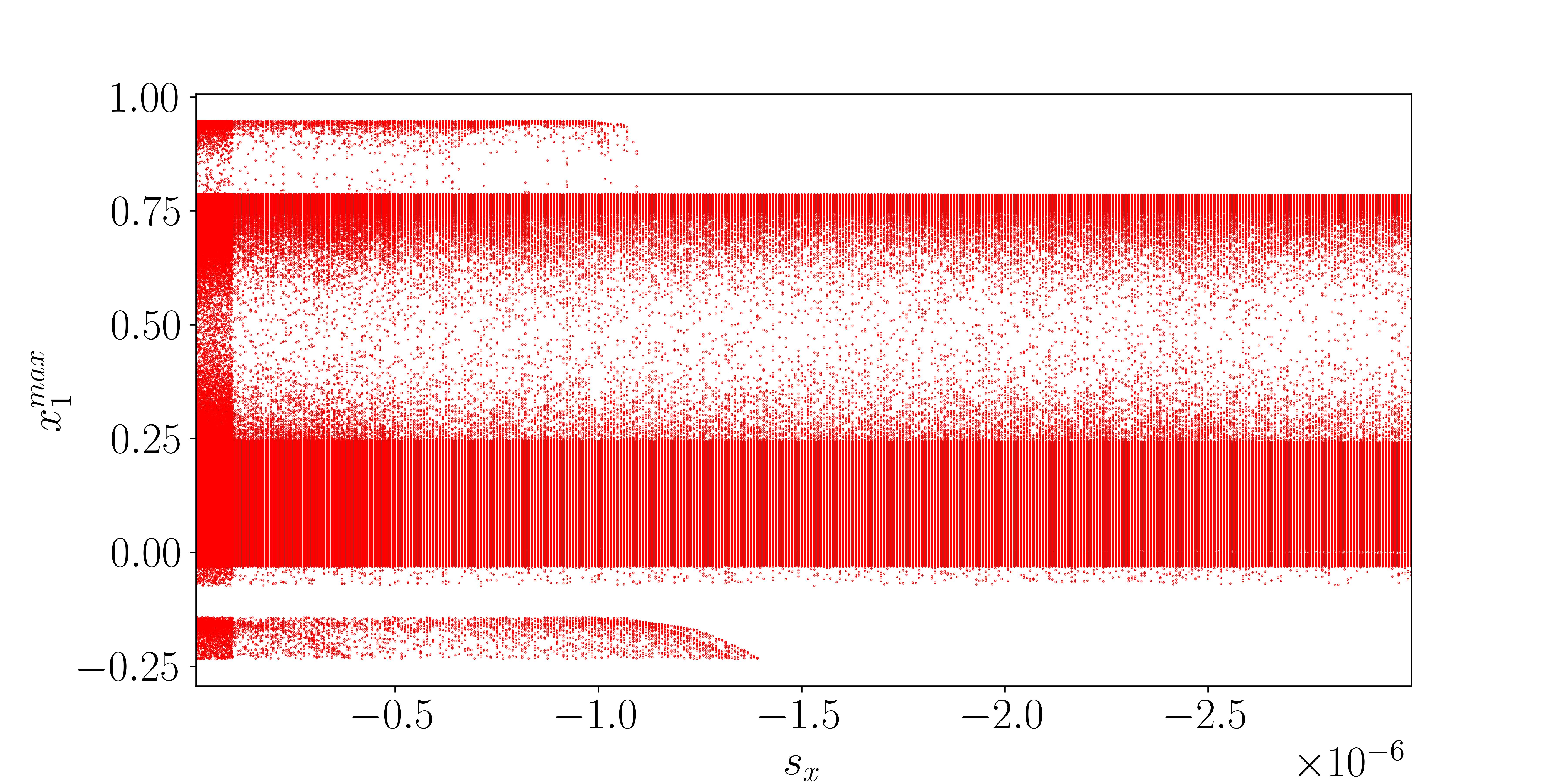}
	\caption{Peaks of the first oscillator in system (2) as a function of constant bias ($s_{x}$) - Bifurcation plot. The contraction of the chaotic attractor occurs at $s_{x}=-1.2\times10^{-6}$. Other parameters are the same as used in Fig.~16.}
	\label{fhn_nc_bif_anx}
\end{figure}
\par In Fig. 17, we plot the bifurcation of the first oscillator in system (2) with the parameter set (C) for varying $s_{x}$. From the figure, we can notice that the size of the chaotic attractor does not change as constant bias is increased. This is stable upto $s_{x} = -1.1 \times 10^{-6}$. When increasing bias to $s_{x} = -1.2 \times 10^{-6}$ the size of the chaotic attractor decreases. This transition indicates the suppression of extreme events and only the chaotic motion is sustained. For better visualization we have shown the chaotic motion for $s_{x}=-1.4\times10^{-6}$ in Fig. 16(j). 
\begin{figure}[!ht]
	\centering
		\includegraphics[width=0.5\textwidth]{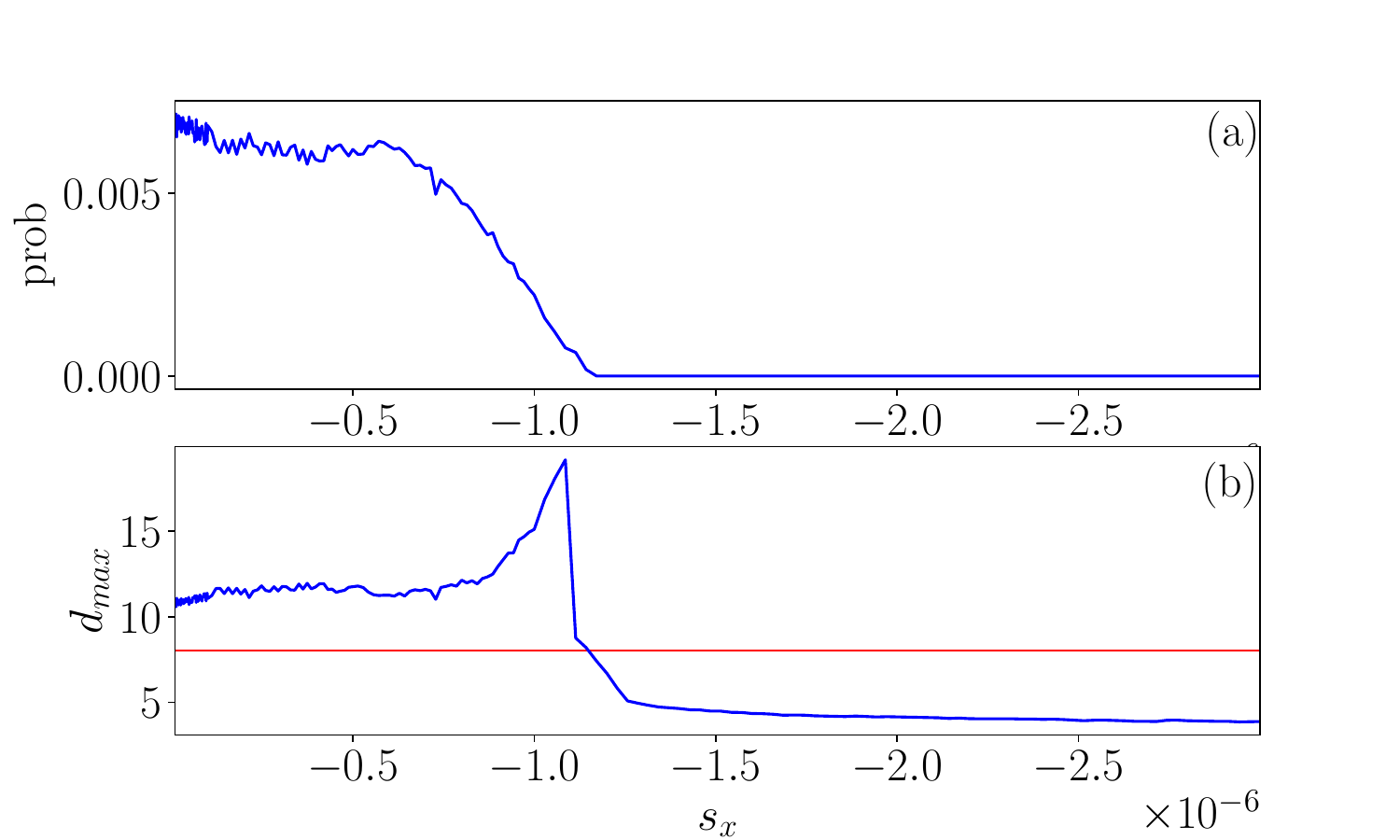}
	\caption{(a) Variation of probability and (b) $d_{max}$ with respect to $s_{x}$ in system (2). The  probability becomes zero and the $d_{max}$ is reduced below $n=8$ line at $s_{x}=-1.2\times10^{-6}$. Other parameters are the same as used in Fig.~16.}
	\label{fhn_nc_dmpb_anx}
\end{figure}
\begin{figure}[!ht]
	\centering
		\includegraphics[width=0.6\textwidth]{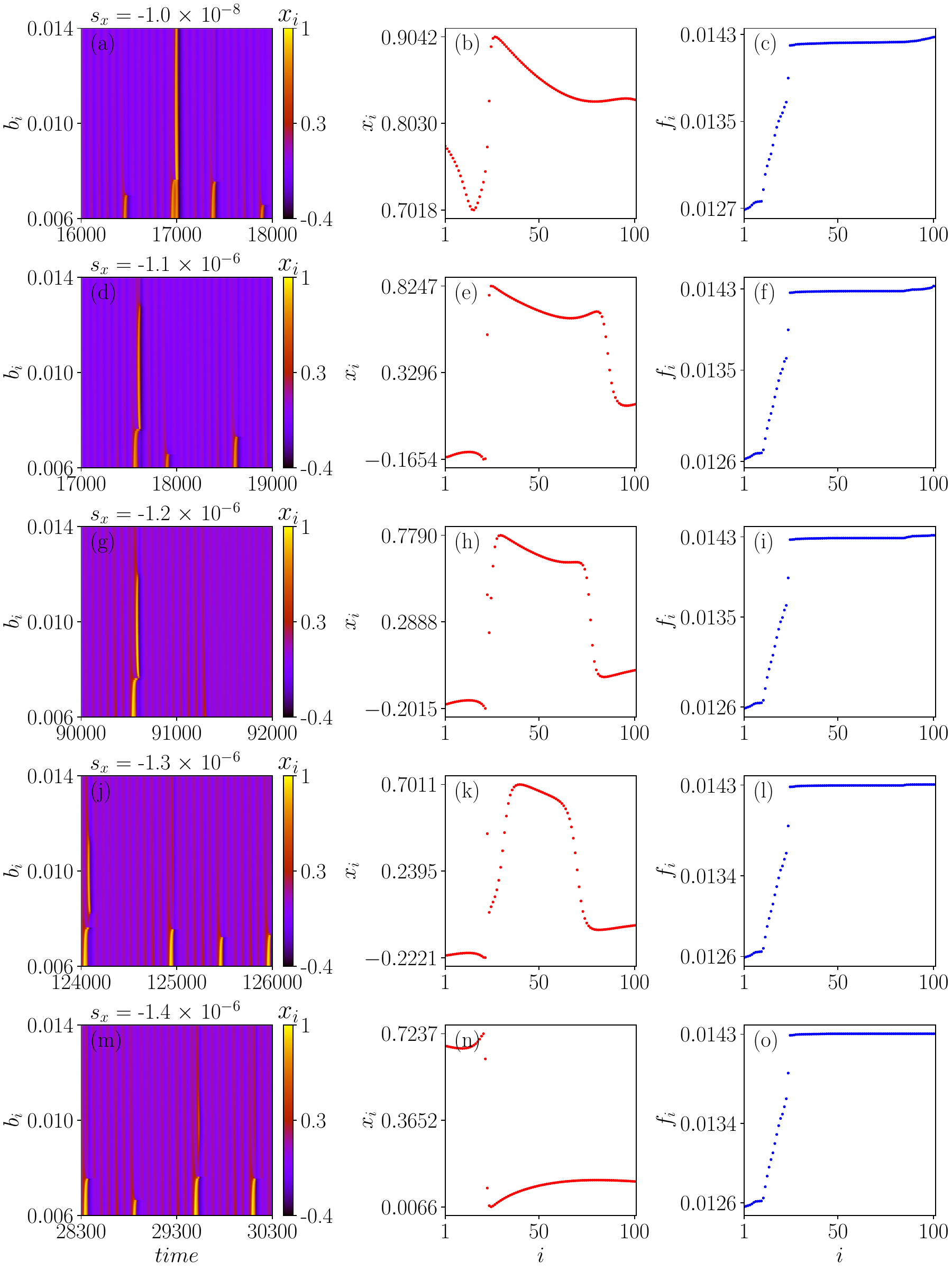}
		\label{fhn_nc_ssf_anx}
	\caption{The columns 1, 2 and 3 represents respectively the spatio-temporal, snapshots and the frequency plots. For all the three columns, from first row to the fifth row, the value of constant bias are $s_{x}=-1.0\times10^{-8}$, $s_{x}=-1.1\times10^{-6}$, $s_{x}=-1.2\times10^{-6}$, $s_{x}=-1.3\times10^{-6}$ and $s_{x}=-1.4\times10^{-6}$ respectively. Other parameters are the same as used in Fig.~16.}
\end{figure}
\par The mitigation of extreme events for varying $s$ is verified from the probability and $d_{max}$ plot and is shown in Fig. 18. The probability of occurence of extreme event becomes zero when the bias value is $s_{x} = -1.2\times10^{-6}$ which can be seen from Fig. 18(a). At this point the value of $d_{max}$ decreases below the threshold line. This can be observed from Fig. 18(b). Further increasing the bias value reduces the $d_{max}$ while the probability remains the same. In Fig. 19, we have depicted the spatio-temporal, snapshot and frequency plots of system (C). The oscillator index is represented in terms of the parameter $b$ in the spatio-temporal plots. The oscillators crossing the threshold value $0.6$ are considered as excited units \cite{r8}. If all the units are in excited state then the extreme event is emanated in the system. Figure 19(a) shows the spatio-temporal plot in which the system exhibits extreme events for the parameter value $s_{x} = -1.0 \times 10^{-8}$. Here, all the oscillators are in the excited state. This can be confirmed from Fig. 19(b) where all the units are crossing the threshold. The same result holds for the parameter values from $s_{x}=-1.0 \times 10^{-7}$ to $-1.0 \times 10^{-6}$. When we increase the bias value to $s_{x}=-1.1 \times 10^{-6}$ the number of excited units gets reduced to 61 and this can be confirmed from the Figs. 19(d) and 19(e), where there are only 61 units crossing the excitability threshold. 
\par Further increasing the bias values to $s_{x}=-1.2 \times 10^{-6}$, $s_{x}=-1.3 \times 10^{-6}$ and $s_{x}=-1.4 \times 10^{-6}$ the number of excited units gets reduced to 47, 29 and 22. These results can be confirmed from the spatio-temporal plots and snapshot plots present in left and middle columns of Fig. 19. The last column in Fig. 19 represents the frequency plots for varying the constant bias. The frequency of the oscillators in N-coupled system is calculated using the formula $f_{i} = \frac{2\pi\gamma_{i}}{\Delta T}$, where $i = 1, 2, 3, ...N$ and $\gamma_{i}$ are the number of maxima in the time series $x_{i}$ of the $i^{th}$ oscillator \cite{r42} and $\Delta T $ refers to time difference between the final $i^{th}$ maxima and the initial $i^{th}$ maxima. This frequency calculation is done for $10^{9}$ iterations. There is no change in the frequency of the oscillators while increasing the bias. Here we note that the constant bias reduces the number of excited units thereby suppressing the extreme events without affecting the frequency of the oscillators. The oscillators from 1 to 10 have the same frequency $f\approx0.0127$. 
\par The oscillators from 11 to 24 have increasing order of frequencies and the oscillators from 25 to 101 has the same frequency $f\approx0.014$ which can be seen from the third column of Fig. 19. The spatio-temporal plots of non-extreme regimes are displayed in Fig. 20. The plots in Fig. 20(a)-(b) correspond to the chaotic regime, where only 22 or lesser number of oscillators are in the excited state (amplitude greater than 0.6). So, there is no possibility for extreme events to occur whereas in Figs. 20(c)-(d) no such protoevents emerge. Only periodic dynamics prevail.
\begin{figure}[!ht]
	\centering
	\includegraphics[width=1.0\textwidth]{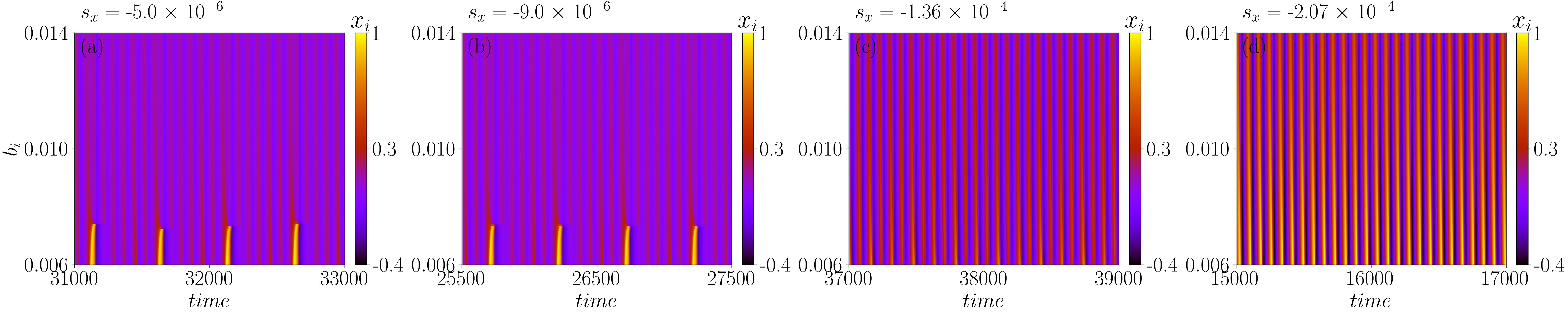}
	\caption{Spatiotemporal plots for non-extreme event regimes for constant bias in membrane potential variable $(x)$. Plots (a) and (b) corresponds to chaotic regimes whereas plots (c) and (d) corresponds to periodic regimes. The values of the constant bias are given on the top of each plot. Other parameters are the same as used in Fig.~16.}
\end{figure}
\subsubsection{Positive constant bias in recovery $(y)$ variable}
\par System (2) with the parameter set (C) exhibits extreme events for the bias value, $s_{y} = 1.0 \times 10^{-8}$. A slight increase in bias, $s_{y} = 2.0 \times 10^{-8}$ decreases the number of large amplitude peaks which can be seen from the Fig. 21(g). The emanation probability of extreme events beyond the threshold line are decreased which can be seen from the Fig. 21(i). Finally, when we increase the bias value to $s_{y} = 2.7 \times 10^{-8}$, the extreme events gets suppressed and only the chaotic motion persists.
\begin{figure}[!ht]
	\centering
	\includegraphics[width=0.6\textwidth]{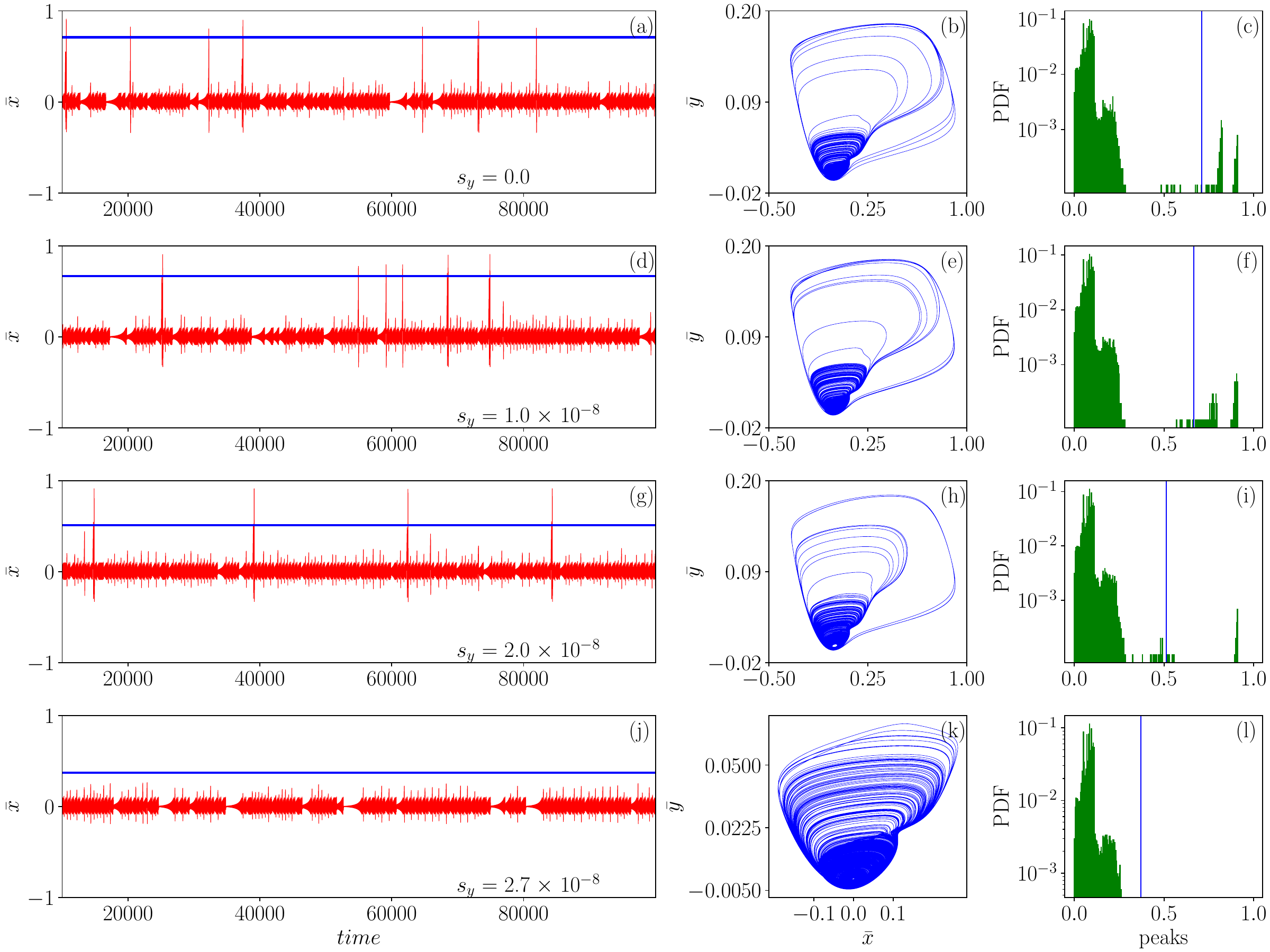}
	\caption{The left, middle and the right columns displays the time evolution, phase portraits and probability distribution plots of system (2). The parameters are $a_{i}=-0.02651$, $c_{i}=0.02$, $k = 0.00128$. Parameter $b$ is distributed from the expression $b_{i} = 0.006 + 0.008(\frac{i - 1}{n - 1})$. From the first row to the fourth row, the value of constant bias are $s_{y}=0.0$, $s_{y} = 1.0 \times 10^{-8}$, $s_{y} = 2.0 \times 10^{-8}$ and $s_{y} = 2.7 \times 10^{-8}$ respectively. The blue line in the first column indicates the threshold.}
	\label{fhn_nc_tspp1_apy}
\end{figure}

\begin{figure}[!ht]
	\centering
		\includegraphics[width=0.5\textwidth]{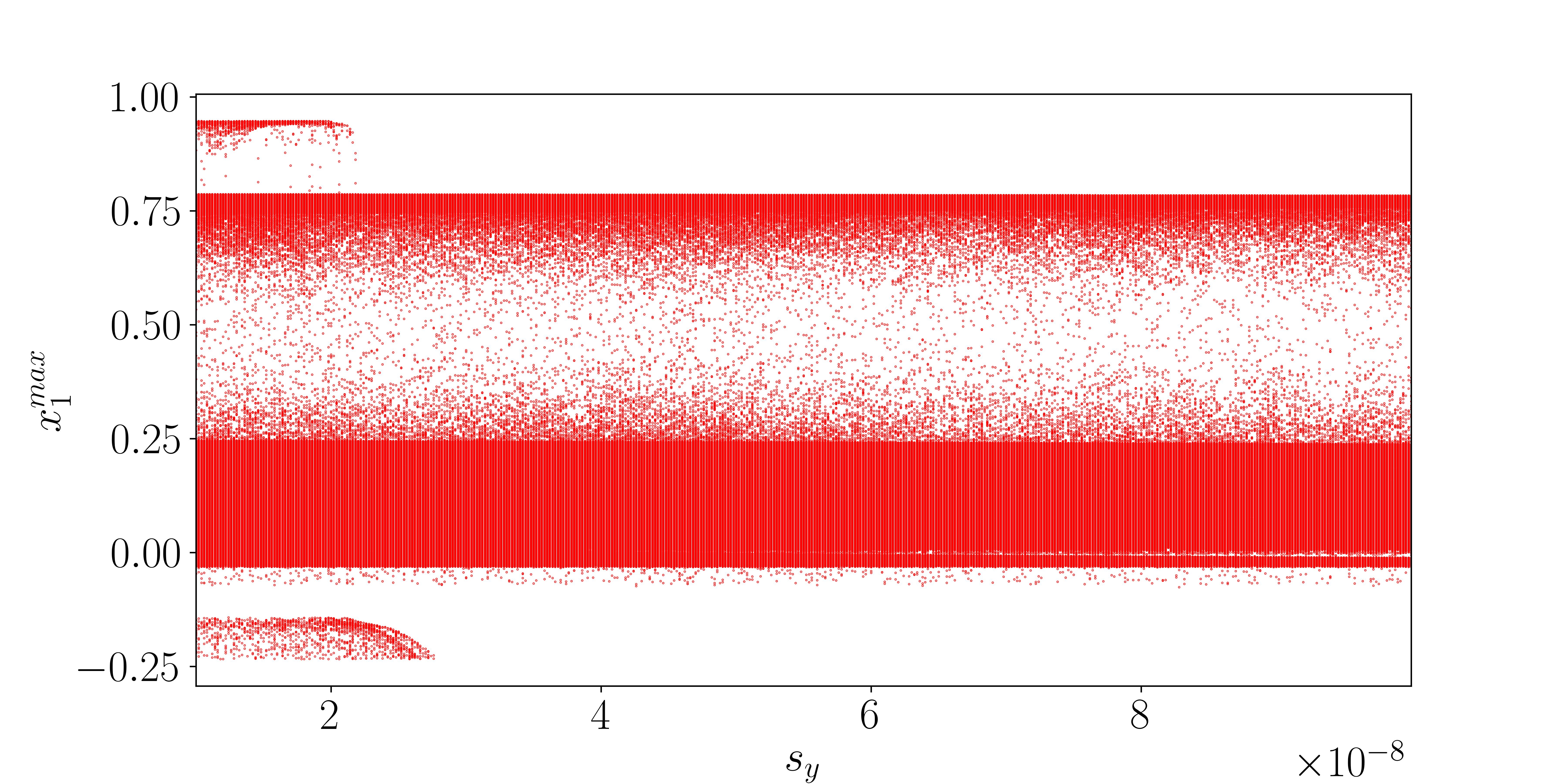}
	\caption{Peak bifurcation diagram of the first oscillator in system (2) showing the contraction of chaotic attractor at $s_{y}=2.4\times10^{-8}$. Other parameters are the same as used in Fig.~21.}
	\label{fhn_nc_bif_apy}
\end{figure} 
\par We have depicted the bifurcation of the first oscillator in system (2) with the parameter set (C) for varying $s_y$ in Fig. 22. There is no change in the size of the attractor upto $s_{y} = 2.3 \times 10^{-8}$. Increasing the value of constant bias to $s _{y} = 2.4 \times 10^{-8}$ initially shrinks the chaotic attractor but after that it has no impact on size of the chaotic attarctor. This shrinkage causes the suppression of extreme events. We have shown clear visualization of this chaotic motion for $s_{x}=2.7\times10^{-8}$ in Fig. 21(j).
\begin{figure}[!ht]
	\centering
		\includegraphics[width=0.5\textwidth]{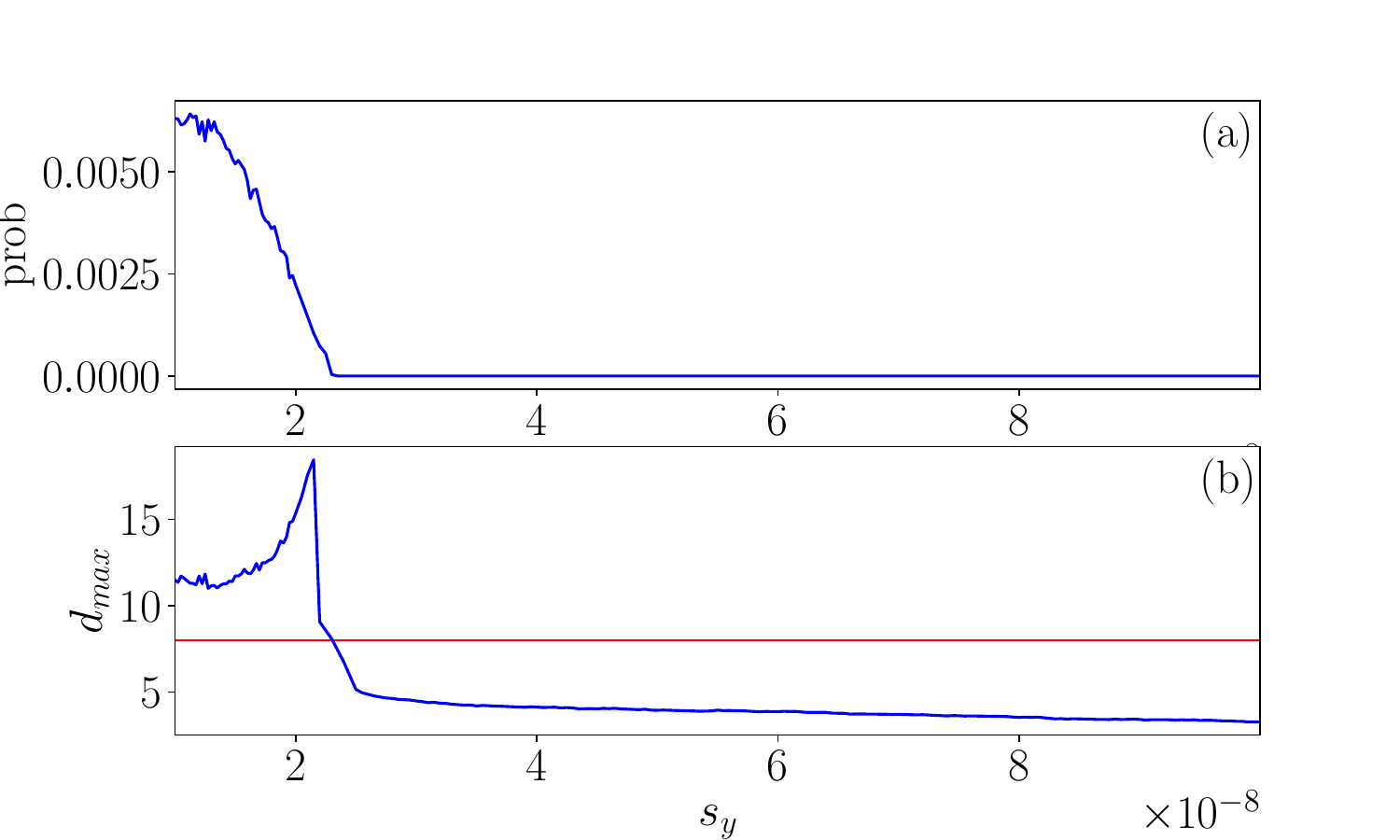}
	\caption{(a) Probability plot  and (b) $d_{max}$ plot varying as a function of $s_{y}$ in system (2). One can observe that the  probability becomes zero and the $d_{max}$ gets reduced below $n=8$ line at $s_{y}=2.4\times10^{-8}$. Other parameters are the same as used in Fig.~21.}
	\label{fhn_nc_dmpb_apy}
\end{figure}
\par The mitigation of extreme events for varying $s$, is verified from $d_{max}$ and probability analysis. When the value bias is $s_{y}= 2.4\times 10^{-8}$, mitigation of extreme events occur and probability becomes zero (Fig. 23(a)). The $d_{max}$ is reduced below the $n$ line (Fig. 23(b)). Further increasing the bias value reduces the $d_{max}$ as the probability becomes zero.
\begin{figure}[!ht]
	\centering
	\includegraphics[width=0.6\textwidth]{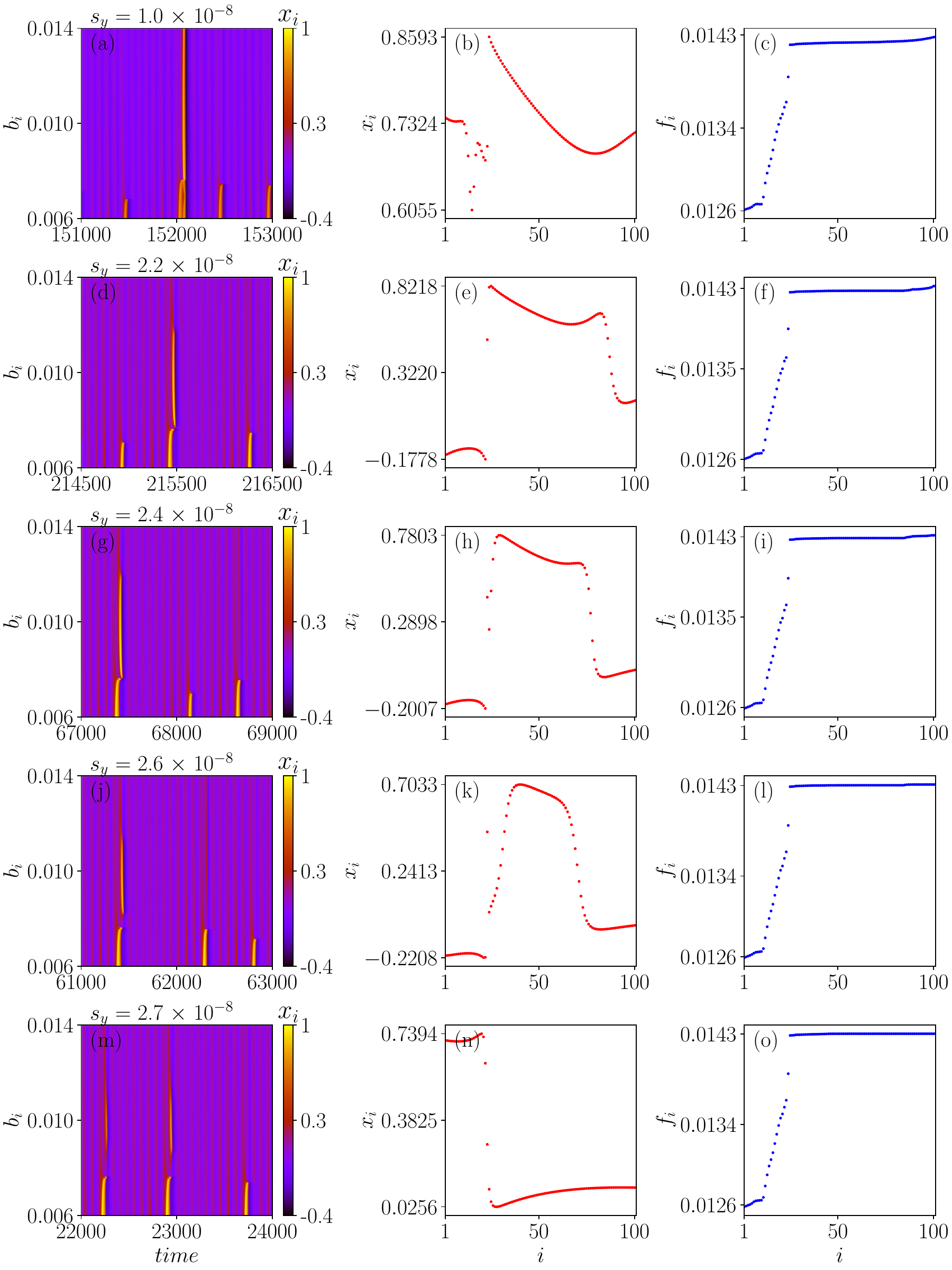}
	\label{fhn_nc_ssf_apy}
	\caption{The spatiotemporal, snapshots and the frequency plots are shown in column 1, 2 and 3 respectively. For all the three columns, from first row to the fifth row, the value of constant bias are $s_{y}=1.0\times10^{-8}$, $s_{y}=2.2\times10^{-8}$, $s_{y}=2.4\times10^{-8}$, $s_{y}=2.6\times10^{-8}$ and $s_{y}=2.7\times10^{-8}$ respectively. Other parameters are the same as used in Fig.~21.}
\end{figure}
\par The spatiotemporal, snapshot and frequency plots of system (C) are displayed in the Fig. 24. The spatiotemporal plot in Fig. 24(a) corresponds to extreme events for the parameter value $s_{y} = 1.0 \times 10^{-8}$. In this regime all the oscillators are in the excited state and this can be confirmed from Fig. 19(b) where all the units are crossing the threshold. When we increase the bias value to $ s_{y}=2.2 \times 10^{-8}$ the number of excited units reduces to 62 which can be confirmed from the Figs. 24(d)  and 24(e). The number of excited units reduces from 62 to 48, 29 and 22 while increasing the bias values to $s_{y}=2.4 \times 10^{-8}$, $s_{y}=2.6 \times 10^{-8}$ and $s_{y}=2.7 \times 10^{-8}$ respectively. These results can be confirmed from the first column of Figs. 24. The frequency plots for varying constant bias values are shown in the last column of Fig. 24. Without disturbing the frequency of the oscillators, the constant bias reduces the number of excited units thereby suppressing the extreme events. The oscillators from 1 to 10 have the same frequency $f\approx0.0126$. The oscillators from 11 to 24 show increasing order of frequencies while the oscillators from 25 to 101 has the same frequency $f\approx0.014$.
\par The spatio-temporal plots for non-extreme regimes have been displayed in Fig. 25. The plots in Fig. 25(a)-(b) corresponds to the chaotic regime, where only 22 or less number of oscillators are in the excited state. Hence there is no possibility for extreme events to occur. Whereas, in Figs. 25(c)-(d), no such protoevents emerge. Only periodic dynamics prevail.
\begin{figure}[!ht]
	\centering
	\includegraphics[width=1.0\textwidth]{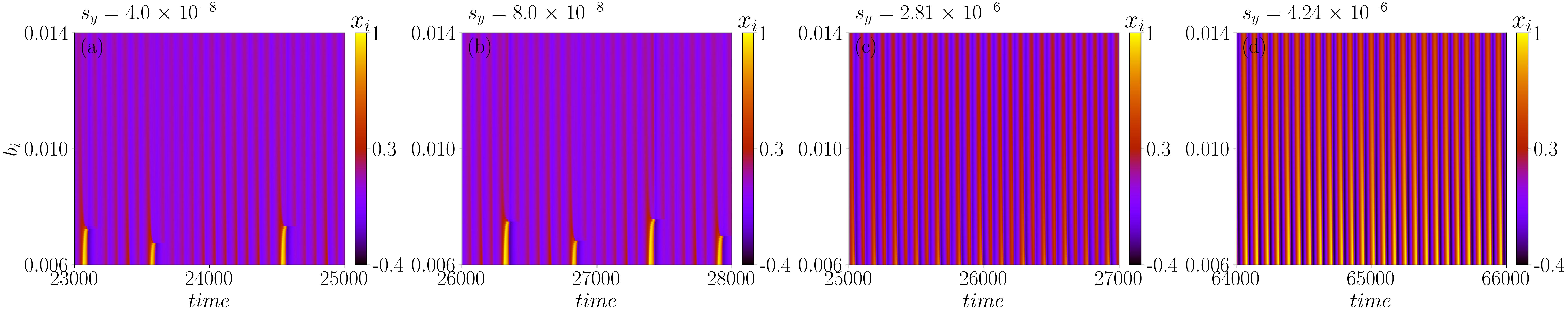}
	\caption{Spatiotemporal plots for non-extreme event regimes. Constant bias is applied in the recovery variable $(y)$. Plots (a) and (b) corresponds to chaotic regimes whereas plots (c) and (d) is for the periodic regimes. The values of the constant bias are given on the top of each plot. Other parameters are the same as used in Fig.~21. }
\end{figure}
\section{Discussion}
\par In the $Two-$coupled system the positioning of manifolds of saddle focus in state space creates a channel-like structure. Through this, the chaotic trajectory escapes and takes a long excursion which is responsible for the emergence of extreme events \cite{r8}. We can see that upon applying a constant bias, the large amplitude oscillation alone decreases in size and number which is discussed in Section 3. And also the long excursion in the phase portrait also decreases and at one point no long excursion is found. This shows that while implementing constant bias, the bias controls the opening and closing of the channel-like structure and completely closes the channel-like structure which prevents the emanation of extreme events. We have also observed that the constant bias has no effect on the phase difference between the two oscillators and is the same when applied to $x$ and $y$ state variables. We have also implemented constant bias in the state variables $x$ in the positive spatial direction and $y$ in the negative spatial direction and observed that the large amplitude peaks increases as we increase the constant bias thereby increasing extreme events.
\begin{table}
	\begin{center}
		\scalebox{1.35}{ 
			\begin{tabular} { |c|c|c|c| }	
				\hline
				\textbf{constant bias ($s_{x}$)} &  \textbf{No. of excited units} &  \textbf{constant bias ($s_{x}$)} &  \textbf{No. of excited units} \\
				\hline
				$-1.0 \times 10^{-8}$ & {101} & $-5.0 \times 10^{-6}$ & {21}\\ \hline
				$-1.0 \times 10^{-7}$ & {101} & $-6.0 \times 10^{-6}$ & {21}\\ \hline
				$-1.0 \times 10^{-6}$ & {101} & $-7.0 \times 10^{-6}$ & {21}\\ \hline
				$-1.1 \times 10^{-6}$ & {61}  & $-8.0 \times 10^{-6}$ & {20}\\ \hline
				$-1.2 \times 10^{-6}$ & {47}  & $-9.0 \times 10^{-6}$ & {20}\\ \hline
				$-1.3 \times 10^{-6}$ & {29}  & $-1.0 \times 10^{-5}$ & {20}\\ \hline
				$-1.4 \times 10^{-6}$ & {22}  & $-2.0 \times 10^{-5}$ & {17}\\ \hline
				$-1.5 \times 10^{-6}$ & {22}  & $-3.0 \times 10^{-5}$ & {14}\\ \hline
				$-1.6 \times 10^{-6}$ & {22}  & $-4.0 \times 10^{-5}$ & {12}\\ \hline
				$-1.7 \times 10^{-6}$ & {22}  & $-5.0 \times 10^{-5}$ & {10}\\ \hline
				$-1.8 \times 10^{-6}$ & {22}  & $-6.0 \times 10^{-5}$ & {8}\\ \hline
				$-1.9 \times 10^{-6}$ & {22}  & $-7.0 \times 10^{-5}$ & {5}\\ \hline
				$-2.0 \times 10^{-6}$ & {22}  & $-8.0 \times 10^{-5}$ & {4}\\ \hline
				$-3.0 \times 10^{-6}$ & {22}  & $-8.0 \times 10^{-5}$ & {2}\\ \hline
				$-4.0 \times 10^{-6}$ & {21}  & $-1.0 \times 10^{-4}$ & {0}\\ \hline
		\end{tabular}}
	\end{center}
	\caption{The number of excited units for various values of constant bias $(s_{x})$ in system (2) with parameter set (C).}
\end{table}
  
\par The emanation of extreme events in $Three-$coupled FHN system is because of the interior crisis. The chaotic attractor suddenly expands at $k=0.064$ which is here referred to as the interior crisis. We have also found that while suppressing the extreme events in three coupled FHN system, the constant bias has no effect on the phase difference and is the same when applied to the $(x)$ and $(y)$ state variables.
\par The FHN system is an excitable system. It has been reported in \cite{r8} that the proto-events occur before the emanation of extreme event and serves as the forerunner for the evolution of extreme event in $N-$coupled FHN system. This proto-event stimulates all other oscillators to excite and generate an extreme event. When we implement bias, it causes a de-excitation of the oscillators participating in the extreme event from the excited state. The gradual de-excitation of number of oscillators from the excited state while increasing the bias value (Table 1). For the first three bias values, there is no de-excitation of oscillators. There is a sudden decrement in the excited units to 61 from 101, to 47 from 61 and to 29 from 47 for the values $s_{x} = -1.1 \times 10^{-6}$, $s_{x} = -1.2 \times 10^{-6}$ and $s_{x} = -1.3 \times 10^{-6}$ respectively. The number of excited units is reduced to 22 and remains the same for the range from $s_{x} = -1.4 \times 10^{-6}$ to $-3.0 \times 10^{-6}$ and there is a decrement in the excited units. Finally, at $s_{x}=-1.0 \times 10^{-4}$ there are no excited units and the system exhibits periodic dynamics.
\par The gradual de-excitation of a number of oscillators from the excited state can be seen from Table 2 when the bias value in the $y$ variable is increased. For the first three bias values there is no de-excitation of oscillators. There is a sudden decrement of excited units to 60 from 101, to 55 from 60, to 48 from 55, to 29 from 48 and to 26 from 29 for the values $s_{y} = 2.2 \times 10^{-8}$, $s_{y} = 2.3 \times 10^{-8}$, $s_{y} = 2.4 \times 10^{-8}$, $s_{y} = 2.5 \times 10^{-8}$ and $s_{y} = 2.6 \times 10^{-8}$ respectively. The number of excited units reduces to 22 and remains the same from $s_{y} = 2.7 \times 10^{-8}$ to $7.0 \times 10^{-8}$ and there is a decrement in the excited units for further increase in $s$. Finally, at $s_{y}=1.0 \times 10^{-5}$ there are no excited units. It can also be found from the frequency plots that the frequency of the oscillators participating in the network formation does not change for the applied constant bias values. But extreme events get suppressed completely. This shows that constant bias suppress extreme events without disturbing the frequency of oscillations in the network. Since collective frequency is responsible for emergent phenomena, the low amplitude constant bias does not alter the network configuration of the system.
\begin{table}
	\begin{center}
		\scalebox{1.35}{ 
			\begin{tabular} { |c|c|c|c| }	
				\hline
				\textbf{constant bias ($s_{y}$)} &  \textbf{No. of excited units} &  \textbf{constant bias ($s_{y}$)} &  \textbf{No. of excited units} \\
				\hline
				$1.0 \times 10^{-8}$ & {101} & $7.0 \times 10^{-8}$ & {22}\\ \hline
				$2.0 \times 10^{-8}$ & {101} & $8.0 \times 10^{-8}$ & {21}\\ \hline
				$2.1 \times 10^{-8}$ & {101} & $9.0 \times 10^{-8}$ & {21}\\ \hline
				$2.2 \times 10^{-8}$ & {60}  & $1.0 \times 10^{-7}$ & {21}\\ \hline
				$2.3 \times 10^{-8}$ & {55}  & $2.0 \times 10^{-7}$ & {20}\\ \hline
				$2.4 \times 10^{-8}$ & {48}  & $3.0 \times 10^{-7}$ & {18}\\ \hline
				$2.5 \times 10^{-8}$ & {29}  & $4.0 \times 10^{-7}$ & {17}\\ \hline
				$2.6 \times 10^{-8}$ & {26}  & $5.0 \times 10^{-7}$ & {15}\\ \hline
				$2.7 \times 10^{-8}$ & {22}  & $6.0 \times 10^{-7}$ & {14}\\ \hline
				$2.8 \times 10^{-8}$ & {22}  & $7.0 \times 10^{-7}$ & {12}\\ \hline
				$2.9 \times 10^{-8}$ & {22}  & $8.0 \times 10^{-7}$ & {12}\\ \hline
				$3.0 \times 10^{-8}$ & {22}  & $9.0 \times 10^{-7}$ & {10}\\ \hline
				$4.0 \times 10^{-8}$ & {22}  & $1.0 \times 10^{-6}$ & {9}\\  \hline
				$5.0 \times 10^{-8}$ & {22}  & $3.0 \times 10^{-6}$ & {0}\\  \hline
				$6.0 \times 10^{-8}$ & {22}  & $1.0 \times 10^{-5}$ & {0}\\  \hline
			\end{tabular}}
	\end{center}
	\caption{The number of excited units for varying the constant bias $(s_{y})$ in system (2) with parameter set (C).}
\end{table}
\par Our study also reveals that extreme events are suppressed with very small bias values. The practicality of attaining such small values is possible with the help of function generator which is used to generate electric signals of various forms in such low scales. Optoelectronic devices such as photodiodes and nano-voltaic devices which convert light energy into electrical energy at micro and nano levels can be also be used. So far in the literature as small as 0.2-0.4 mA has been utilized in DBS to treat Temporal Lobe Epilepsy \cite{r47}. But it requires a lot of research to device such instruments that would produce signals in such low ranges with high precision. Though it is difficult to attain at the present stage, with evolving technology, it will be acheived soon.
\section{Conclusion}
\par In this work we have studied the influence of constant bias on extreme events in $Two$, $Three$ and $N-$coupled FHN system. We observed the emergence of extreme events in three coupled FHN system and found that interior crisis is the mechanism behind the  emergence of extreme events. We found that the extreme events in $Two$, $Three-$ and $N-$coupled FHN neuron system gets mitigated through shrinkage of chaotic attractor and the chaotic attractor transits into periodic attractor through reverse period-doubling bifurcation. We have illustrated and verified the mitigation of extreme events using $d_{max}$ and probability plots. We have also shown the structure behind the mitigation of extreme events using the spatiotemporal plot and have shown that the number of excited units decreases in $N-$coupled FHN system while varying the bias. In addition to that, we have demonstrated that constant bias suppresses extreme events without the changing the frequency of the $N-$coupled FHN system. Applying constant bias does not directly alter the number of protoevents, rather it gradually affects the excitability of the units participating in the extreme events thereby reducing the number of excited units. One of our main observation is that the network configuration of the  system does not change during the suppression of extreme events. Further, the approach used in our study has the potential to treat not only epilepsy but also various kinds of neuronal disorders like Parkinson disease, tremors, dystonia and tourette syndrome. These are some of the long-term significant potential applications of this study. Developing more tools to mitigate these extreme events is still underway.
\section*{Acknowledgement}
RS thanks Bharathidasan University for providing, University Research Fellowship (URF). RS also thanks Prof. P. Muruganandam, Department of Physics, Bharathidasan University, for his numerical support in constructing RKF45 adaptive algorithm. SS wishes to thank DST-SERB, Government of India for providing National Post-Doctoral Fellowship under Grant No. PDF/2022/001760. The work of MS forms a part of a research project sponsored by Council of Scientific and Industrial Research (CSIR) under the Grant No. 03/1482/2023/EMR-II.
\section*{Authors Contributions}
All the authors contributed equally to the preparation of manuscript.
\section*{Data Availability Statement}
The data that support the findings of this study are available within this article. 

\end{document}